%% file: oatao.tex
\documentclass{article}

\usepackage{pgf, pgfarrows, pgfplots}
\usepackage{multirow}
\usepackage{url}
\usepackage{footnote}
\makesavenoteenv{tabular}
\makesavenoteenv{table}
\usetikzlibrary{arrows,shadows,positioning}
\usepgfplotslibrary{external}

\usepackage[left=3cm,right=4cm,top=2cm,bottom=3cm]{geometry}

\newcommand{\varLinuxKernel}{4.15.0-29-generic}

\newcommand{\varChromeVersion}{67.0.3396.99}

\newcommand{\varNbTestUnits}{40}

\begin{document}

\title{Google QUIC performance over a public SATCOM access}

\author{Ludovic Thomas$^*$, Emmanuel Dubois$^{\dag}$, Nicolas Kuhn$^{\dag}$ , Emmanuel Lochin$^*$
~\\
~\\
$^*$ ISAE-SUPAERO, Toulouse, France
~\\
$^{\dag}$ CNES, Toulouse, France
}

\date{}

\maketitle

\begin{abstract}
Google QUIC accounts for almost $10$\,\% of the Internet traffic and the protocol is not standardized at the IETF yet. We distinguish Google QUIC (GQUIC) and IETF QUIC (IQUIC) since there may be differences between the two. 
Both Google and IETF versions run over UDP and cannot be split the way satellite systems usually do with TCP connections. 
The need for adapting any-QUIC parameters needs to be evaluated.
Since GQUIC is available, we analyze its behavior over a satellite communication system. 
In our evaluations, GQUIC quick connection establishment does not compensate an inappropriate congestion control.
The resulting page downloading time doubles when using GQUIC as opposed to the performance with optimized split TCP connections.
This paper concludes that specific tuning are required when any-QUIC runs over a high BDP network.
\end{abstract}


\input{introduction}
\input{testbed}

\input{results}

\input{related_work}

\input{discussion}

\bibliographystyle{plain}
\bibliography{oatao}

\end{document}

%% file: introduction.tex
\section{Introduction}
\label{sec:introduction}

Quick UDP Internet Connections (QUIC) is a transport-layer protocol running on top of User Datagram Protocol (UDP)~\cite{quic-deploy-doc} developed by Google since 2012 and currently under discussion at Internet Engineering Task Force (IETF)~\cite{quic-wg}. We distinguish Google QUIC (GQUIC) and IETF QUIC (IQUIC). At the time of writing (December 2018), most of the milestones that are related to a first specification of IQUIC (also called QUIC v1) are expected in July 2019 while extansions on \textit{e.g.} multipath IQUIC are expected in 2019 and 2020. Even if IQUIC is not standardized yet, Google has deployed services over QUIC: GQUIC accounts for $2.6$\,\% to $9.1$\,\% of the Internet traffic with rapidly changing versions~\cite{quic-in-the-wild}. To contribute to the specifications of the protocol at IETF, we need to assess the performance of the actually deployed version to justify the need for modifications for high Bandwidth Delay Product (BDP) networks. 

Any-QUIC benefits from years of development, rapid deployment and large scale testing. Despite evolving versions, several properties are expected to remain invariant such as encryption of both application data and transport parameters. 
Fully-encrypted any-QUIC might lead to discrepancies between Over-The-Top (OTT) protocol design choices and Internet Service Provider (ISP) policing mechanisms.
While OTT and ISP may not always be seen as competitors~\cite{ott-isp}, the deployment of any-QUIC could lead to some ISP issues: (1) to select the appropriate Quality-of-Service (QoS) policy for the applications carried out; (2) to enable the right shaping policy according to both end user's contracts and access network characteristics; or (3) to optimize the use of the constrained resources such as on cellular networks.

As a matter of fact, operational ISP networks do not evolve at the same pace as End-to-End (E2E) protocols. Furthermore, they should not only be influenced by new emerging protocols, but also with existing and potentially old fashioned protocol stacks. Indeed, both the low Transmission Control Protocol (TCP) Initial congestion Window (IW) values measured in~\cite{iw-status} and the analysis of TCP variants in the wild~\cite{tcp-variants-in-wild} highlight that some web services are still using outdated transport protocol flavours. 
Any-QUIC could balance the part of old stacks currently used which is a great illustration of the impact OTT have over Internet traffic. 

SATellite COMmunication (SATCOM) networks typically break the TCP connections to adapt the transport protocol for long delay links. 
Although recent E2E protocols may exhibit decent performance over high BDP paths, splitting TCP allows for adaptation of both TCP slow-start and loss-recovery mechanisms. This results in lower page load times~\cite{pep-satcom}. Moreover, older stacks would anyway need specific acceleration. It is worth pointing out that cellular networks may also introduce the same kind of Performance Enhancing Proxy (PEP) to adapt \textit{e.g.} TCP for the upcoming Fifth-Generation Mobile Communications System (5G): this is not only seen through research papers~\cite{pep-5g-1,pep-5g-2} but also in 3rd Generation Partnership Project (3GPP) study items~\cite{tcp-3gpp}. In this context, the trend towards the deployment of protocols like any-QUIC questions the actual E2E protocols' adaptations. This motivates this study that assesses the performance obtained by GQUIC when it comes to downloading two different web pages over a geostationnary satellite.

To the best of our knowledge, this paper reports the first evaluations of GQUIC using a real public SATCOM access by assessing the web browsing QoE. Our main findings are: 
\begin{itemize}
	\item for a large web page, the page load time is approximately twice longer with GQUIC compared to Transport Layer Security (TLS)/TCP (section \ref{sec:results:plt});
	\item this difference in larger page load time resides in the poor performance of the non-delegated Congestion Controller (CC) in GQUIC (section~\ref{sec:results:cc});
	\item although faster, GQUIC connection establishment does not compensate the above issue (section \ref{sec:results:handshake}).
\end{itemize}

%% file: testbed.tex
\section{Experiment setup}
\label{sec:testbed}

We exploit a public SATCOM Internet access and repeatedly download two
pages with different profiles. 
We report raw Quality-of-Experience (QoE) metrics, such as Page Load Time (PLT), that represent today's end user experience.
We do not pretend to properly assess the QoE of web browsing since it depends a lot on the page that is downloaded.
We mainly report trends on some QoE-related web browsing parameters.

Our approach provides a 
fair comparison between GQUIC and a SATCOM-optimized TCP.
Controlled experiments could hardly be envisioned since the operator's ground
segment implements specific optimizations and IQUIC available frameworks may
not be relevant~\cite{long-look}.

For the sake of reproducible science, we have released the scripts that have been used to generate our results~\cite{quxa-public-repo}.

\subsection{SATCOM and 4G Internet accesses}
\label{sec:testbed:satellite}

To better explain the behavior of GQUIC over a SATCOM access, we also perform some tests with a 4G access as a reference. This section focuses on the description of the SATCOM Internet access. 

The public SATCOM operates in Europe with geostationary
satellites. We have used a KA-SAT PRO25Go access. 
To roughly estimate the likely performance of this access and
provide an initial sanity check, 
we have measured that the network is less congested between $2$\,pm and $4$\,pm
and have decided to run our evaluations at that moment of the day. 
Early evaluations have showed that TCP is split both at the gateway and at the terminal\footnote{TCP PEP deployement has been brought to light based on: (a) computations of TCP-level and TLS-level Round Trip Time (RTT); (b) TCP sequence number evolution compared to expected slow start profile and (c) the \texttt{traceroute} tool.}. However, we can not have much more information on the stacks that are used within the operator network or at the terminal. 
 
The data plan of our contract limits the variety of performed experiments, but our tests let us assess the estimated SATCOM end-users' QoE. Furthermore, we reckon this also illustrates the impact of OTT protocol design decisions over a SATCOM provider. 

A description of a generic SATCOM architecture can be found
in \cite{satcloudran, nc-satcom}. 

\subsection{Web pages and QoE}
\label{sec:testbed:metrics}
To assess the QoE of a web browsing user, we measure the Page Load Time (PLT), defined
as the elapsed time between the \texttt{connectStart} and \texttt{loadEventEnd}
events \cite{navigation-timing}. For an extensive assessment, other metrics could have been considered such as the time to render or the time to first paint. However, the relevance of these metrics depends on the page that is downloaded and we could not consider many web pages characteristics since the pages needed to be available \textit{via} GQUIC. Our objective is to identify trends and high level comparison of the two protocol stacks.

\begin{figure}
	\centering
	\includegraphics[width = 0.7\linewidth]{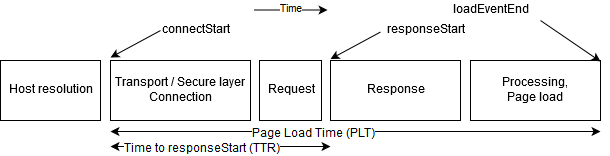}
	\caption{Steps of a page load~\cite{navigation-timing}}
	\label{fig:testbench:metrics}
\end{figure}

As shown in Figure \ref{fig:testbench:metrics}, the PLT can be decomposed into (1) the time needed to complete the handshake and send the request and (2) the time required to
download the content and to process it. In addition to the PLT, we measure the Time To \texttt{responseStart} (TTR), the moment at which the user receives the first byte of the Hypertext Transfer Protocol (HTTP) response from the server. This helps us assess the contribution of the connection establishment and the request
transmission in PLT. Host resolution is not considered. Metrics are measured when accessing two different targets : 
\begin{itemize}
	\item \emph{Target A :} one picture with a $5.3$\,MB total size; 
	\item \emph{Target B :} one Google's 404 page with two objects and $11$\,kB of total size.
\end{itemize} 

Both page are hosted on Google's GQUIC capable servers. 
We acknowledge that selecting these targets has an impact on the relevance of our conclusions. We do not aim at assessing the performance of GQUIC for any page download but rather expect to compare its behaviour when it comes to downloading large web objects. 


\subsection{Scenario configuration}
\label{sec:testbed:client}

Involved protocols collect network and server information using parameters caching, TCP Fast Open (TFO)~\cite{tfo}, GQUIC discovery~\cite{quic-discovery} and GQUIC connection resumption\footnote{Also called zero-RTT and denoted 0RTT in the following.} mechanisms. The cached parameters are then used to improve the following loads\footnote{For instance: caching the certificates reduces the handshake duration both with GQUIC and with TCP/TLS; caching the RTT estimate prevents spurious retransmissions during the GQUIC handshake; etc.}. To analyze their impact on PLT and TTR, each test unit is composed of three web pages downloads before purging the browser profile. For each load, the client fetches one of the web pages and then closes the browser when the page is retrieved. Elapsed time between two loads is uniformly distributed between $5$ and $15$ seconds. 

Automated weather and link quality reports are linked to the test units. Measures were performed during good weather conditions, with no rain and few clouds. Average link characteristics are presented in Table~\ref{tab:qos}. We also provide worth-noting comparison of the size of the targets with the BDP and the IW in Table~\ref{tab:odg}. We assume servers' default IW is set to $32$ TCP Maximum Segment Size (MSS) as observed in GQUIC source code~\cite{chromium-source-code}. The actual IW may not be exactly 32, but the main point of this discussion is to highlight the differences between the size of the file and the expected IW. We use the Selenium automation tools to control the browser and retrieve PLT and TTR. Tests are operated on a laptop with \varLinuxKernel{} Linux.

\begin{table}
	\centering
	\caption{SATCOM and 4G access networks}
	\label{tab:qos}

	\begin{tabular}{c|c|c}
		Metric & SATCOM & 4G \\
		\hline
		\hline
		Downlink capacity & $25$\,Mbps & $5$\,Mbps \\
		Uplink capacity & $5$\,Mbps & $2.5$\,Mbps \\
		RTT & $750$\,ms & $80$\,ms \\
	\end{tabular}
\end{table}

\begin{table}
	\centering
	\caption{Ratios between target sizes, BDP and IW - \textit{e.g.} Target A is $113$ times larger than the IW}
	\label{tab:odg}
	\begin{tabular}{c|c|c|c}
		\multirow{2}{*}{Target} & \multicolumn{2}{c|}{Size / BDP} & \multirow{2}{*}{Size / IW}\\
		& SATCOM & 4G & \\
		\hline
		\hline
		A ($5.3$\,MB) & 4.8 & 212 & 113\\
		B ($11$\,kB) & 0.01 & 0.44 & 0.2\\
	\end{tabular}
\end{table}

\subsection{Few words on the browsers}
\label{sec:testbed:browser}

An analysis of \texttt{Chrome}'s behavior combined with~\cite{quic-deploy-doc, quic-discovery} has provided us with the following expectations:
\begin{itemize}
	\item When \texttt{Chrome} starts, it opens connections with several servers. If GQUIC is enabled, we have identified that \texttt{Chrome} benefits from those loads to open several GQUIC connections. To ensure the same start point whether GQUIC is enabled or not, we have decided to block any GQUIC traffic not intended to the server holding the web-page or its objects. We have done that since we do not want \texttt{Chrome}'s specific implementation of \texttt{Chromium} to have an impact on the conclusions that are proposed in this article. 
	\item Before using GQUIC, \texttt{Chrome} needs to learn about its availability. GQUIC discovery procedure is described in more details in~\cite{quic-discovery, quic-deploy-doc}. Important to note here is that \texttt{Chrome} always use TCP for the first time it contacts a server.
	\item Since \texttt{Chrome} is restarted between each load, it will neither use GQUIC nor TLS connection resumption. This is due to the way \texttt{Chrome} deals with certificates, not to any limitation in the two protocols. We thus expect only 1RTT GQUIC connections.  
\end{itemize}

For each test unit, we use two instances of \texttt{Google Chrome \varChromeVersion{}}:
\begin{itemize}
	\item \texttt{ChromeQuic}: GQUIC is enabled with Bottleneck Bandwidth and Round-trip propagation time (BBR) congestion control instead of CUBIC\footnote{This is performed following the method described in~\cite{enable-quic-bbr-chrome}. The objective is to get a consistent comparison as BBR is expected to be deployed for TCP on Google's infrastructures~\cite{bbr-on-google}.}. We request BBR on the server side \textit{via} flags during the handshake, but cannot assess the BBR version used by Google servers. However, we can expect that Google had deployed the last version at the time of testing.  
 	\item \texttt{ChromeNoQuic}: GQUIC is disabled. 
\end{itemize}

\texttt{ChromeNoQuic} will always use a HTTP2-TLS1.2-(split)TCP (HTT) stack, whereas \texttt{ChromeQuic} starts with HTT and switches to HTTP2-GQUICv39-UDP (HQU) whenever possible.   
For both instances, TFO is enabled and content caching is disabled. Changes described in this section are the only ones performed on publicly available \texttt{Google Chrome}.

%% file: results.tex
\section{Results}
\label{sec:results}

\subsection{PLT for a large page}
\label{sec:results:plt}
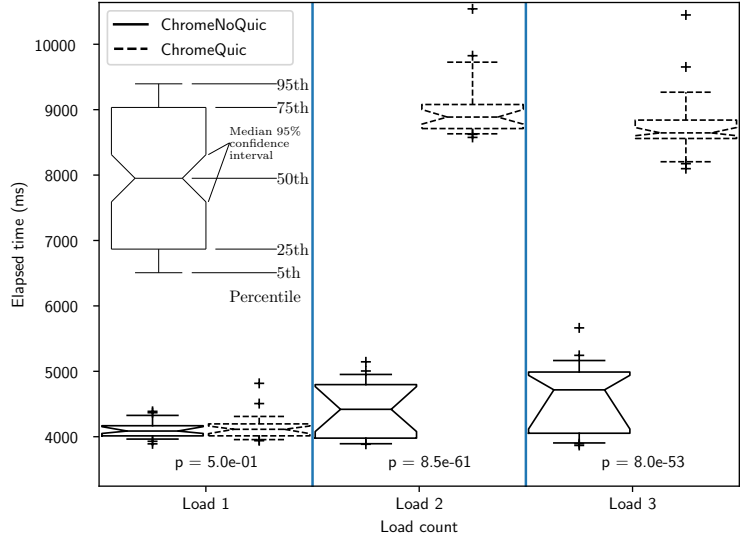
\begin{figure}
	\centering
	\resizebox{0.7\linewidth}{!}{\input{Figure_paper_PLT_A.pgf}}
	\caption{PLT for target A (large picture) on a SATCOM access}
	\label{fig:results:plt:target_a}
\end{figure}

We denote by HTT the HTTP2-TLS1.2-(split)TCP stack and by HQU the HTTP2-GQUICv39-UDP stack.
Figure~\ref{fig:results:plt:target_a} presents the PLT for target A as a function of the load index since last profile purge. Metrics are computed according to~\cite{leboudec} over \varNbTestUnits{} test units. At the bottom of the figure, we also provide the \emph{Welch's t-test} $p$ parameter~\cite{welch} of the two distributions under the \emph{null} hypothesis. For the first load, we expect similar performance for both browser versions as they use the same protocol stack (HTT). The two distributions are indeed located at the same time values and they both show low dispersion. However, with \texttt{ChromeQuic} learning about GQUIC availability, HQU is then permanently used for that browser starting from the second load. 

We observe the worst performance with HQU where PLT is up to twice longer than \texttt{ChromeNoQuic} which uses HTT. The extremely low $p$ values confirm the statistical relevancy of that observation. Each of \texttt{ChromeNoQuic} and \texttt{ChromeQuic} exhibits a higher PLT dispersion and an intersection of the confidence intervals for loads 2 and 3. It substantiates that above mentioned optimization mechanisms are all performed within one load and we do not expect any further evolution of the PLT with additional loads.

\subsection{Packet sequence numbers rate}
\label{sec:results:cc}
\begin{figure}
	\centering
	\includegraphics[width=0.7\linewidth]{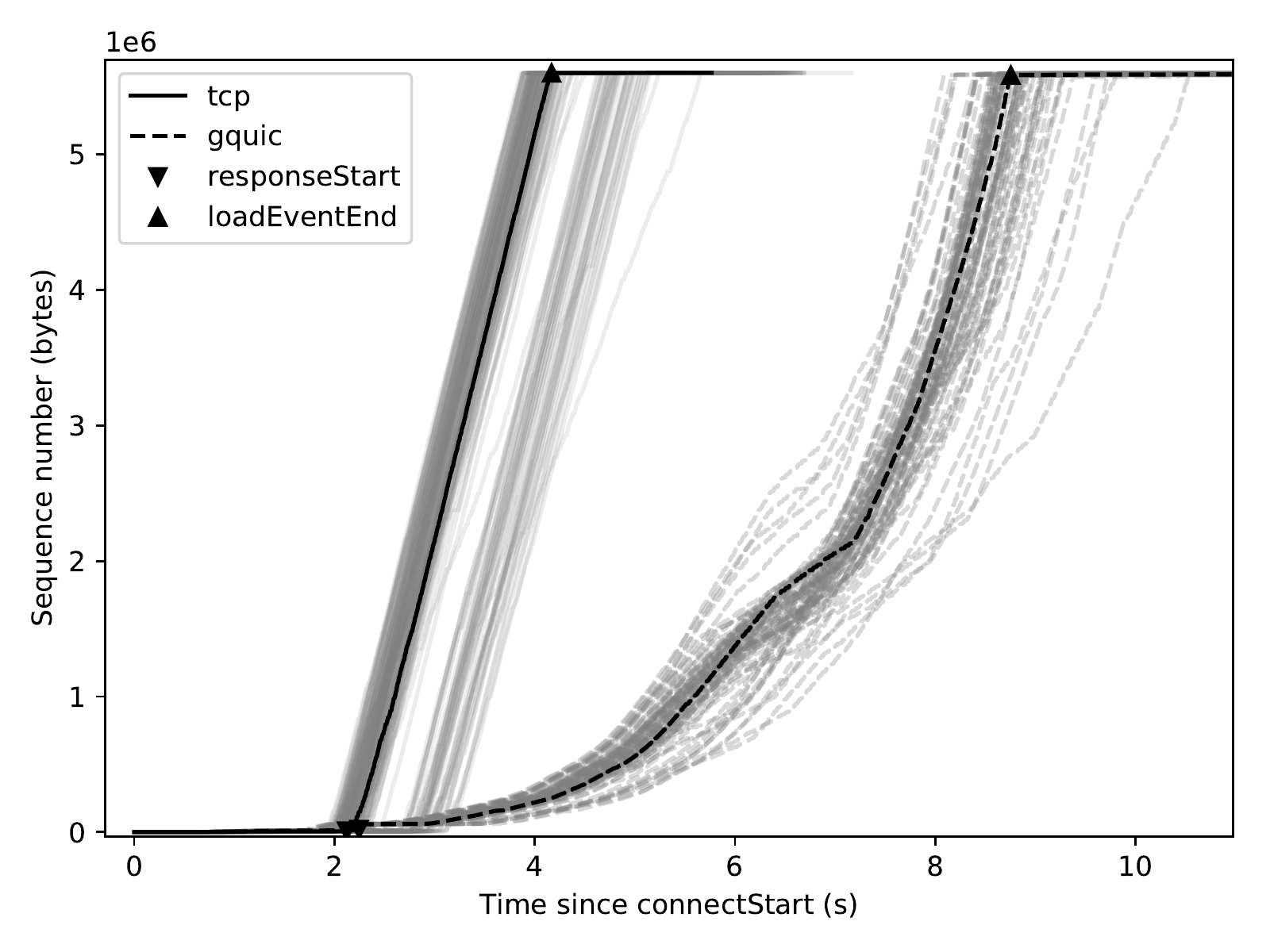}
	\caption{Evolution of the sequence number for the flows downloading the target A on a SATCOM access}
	\label{fig:results:seq_numbers}
\end{figure}
\begin{figure}
	\centering
	\includegraphics[width=0.7\linewidth]{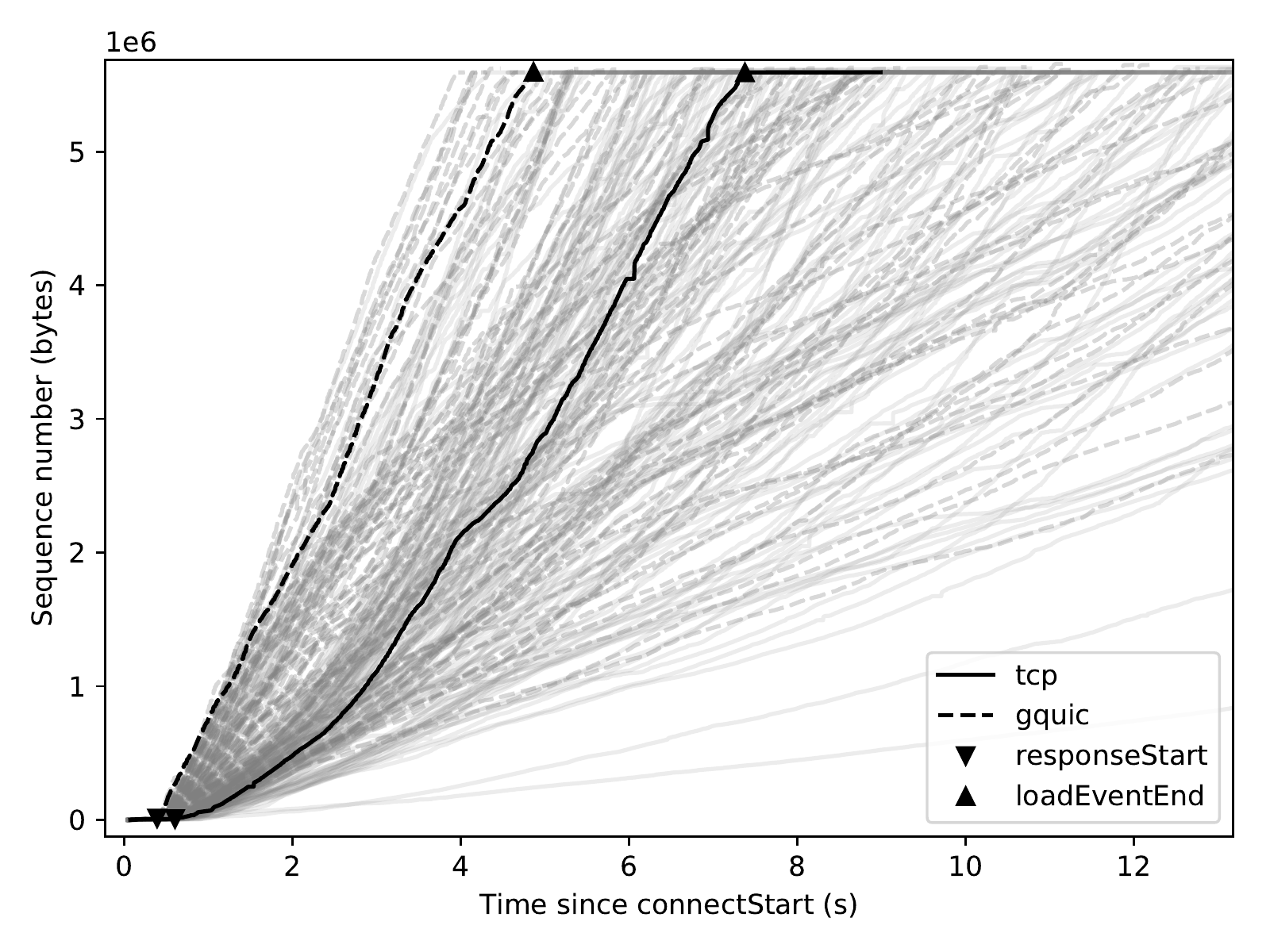}
	\caption{Evolution of the sequence number for the flows downloading the target A on a 4G access network}
	\label{fig:results:seq_numbers_4g}
\end{figure}

To understand the performance gap between HQU and HTT, we first focus on the second load of target A and more particularly on the second phase of a PLT, \textit{i.e.}, the time elapsed between the reception of the first response byte and the reception of its last. 

To mitigate the issue of encryption keys, we define a sequence number equivalent for GQUIC connections. It is defined as the cumulative sum of the bytes received over the connection, scaled in order to reach the same last value as with TCP. We recognize that the value may present local differences compared to stream offsets embedded in GQUIC packets. Nonetheless, we do believe that global behaviors can be compared.

Figure \ref{fig:results:seq_numbers} presents those computations on a subset of HTT and HQU connections. Downward [resp. upward] triangles report the location of TTR [resp. PLT] measurements. We observe that the  HQU stack fires the \texttt{responseStart} event before HTT. However, the download is completed way after. This can be explained by HTT getting ``up-to-speed'' and showing a stable and high goodput, while HQU ramps up its transmission rate slowly.  

We first focus on HQU. To discriminate the origin of this slow increase between (a) a UDP throughput control from the ISP or (b) the BBR Congestion Controller (CC) itself, we can compute the duration of the BBR \emph{Startup} phase\footnote{Note that BBR \emph{Startup} phase uses the same exponential increase profile as Slow Start or Hystart and only differs in the exit condition.} : $t_s = \ln_2(\mathcal{B}/\mathcal{W}_i)\mathcal{R}$, with $\mathcal{B}$ the BDP of the link, $\mathcal{R}$ its RTT and $\mathcal{W}_i$ the IW. For the given parameters of the whole E2E link\footnote{Measured throughput : $25$Mbps. Measured RTT : $750$ms. Default IW : 32.} we obtain $t_s \approx 3.5$s. Compared to expected behavior of BBR we can note on Figure \ref{fig:results:seq_numbers} a \emph{Startup} phase of approximatively $4$s (between $2$s and $6$s), followed by three constant-rate segments. The observation is in line with the computed value. With target A only four times bigger than the BDP (Table \ref{tab:odg}), we can note that the CC above GQUIC spends more than two third of the download in its \emph{Startup} phase. It means that we can expect a similar performance for any other CC using a binary search including TCP New Reno and CUBIC, as long as they are implemented above GQUIC. 

In comparison, HTT achieves a near-constant downloading rate, as noted by the quasi-linear increase of sequence numbers. As explained in Section \ref{sec:testbed:satellite}, the TCP path is split into three connections. One for the satellite link and one at each of its edges: connecting the server to the GateWay (GW) and the Satellite Terminal (ST) to the client (see Figure \ref{fig:results:handshake}). Let's suppose the segment $PEP_{GW}$ $\rightarrow$ $PEP_{ST}$ uses proprietary protocols and does not require any startup probing phase. The TCP slow start can be neglected for the segment $PEP_{ST}$ $\rightarrow$ Client since its RTT is around $1$\,ms. Finally, the binary search is also expected to last less than $1$ RTT for the segment Server $\rightarrow$ $PEP_{GW}$ before the later toggle down the emission with its flow control\footnote{Throughput before reaching flow control limitation : $25$Mbps. Computed RTT based on the difference of the RTT with the Server and with the GW : $30$\,ms. Default IW : 32.}. Computations are again in line with the observed values as we note that the final constant TCP throughput is reached in less than $50$ms on the row data of Figure \ref{fig:results:seq_numbers}. In our scenario, splitting TCP and using proprietary protocols in the central segment allows each outer segment to present low BDP and thus permit a fast binary search. On the contrary, GQUIC cannot be split because of transport-level encryption. 

To better explain the poor performances of GQUIC, we run the same computations on a comparative 4G access link. Results are shown in Figure~\ref{fig:results:seq_numbers_4g}. First, we note the HQU stack fires the \texttt{responseStart} event before HTT and the median gap is around $90$ms. Section \ref{sec:results:handshake} provides insights for that difference. Second, we can note that the BDP of the path is here lower than the IW. Thus, the \emph{Startup} phase is completed in less than $1$ RTT. And last, the CCs spend the most part of the load in their \emph{Steady} state because target A is significantly bigger than the BDP (Table \ref{tab:odg}). On that state, GQUIC shows better performances which is consistent with studies performed on non-split paths~\cite{long-look}. On the geostationary link, the gap in TTR was weak and GQUIC was penalized in its slow start compared to split TCP.

\subsection{Focus on the handshake}
\label{sec:results:handshake}
\begin{figure}
	\centering
	\includegraphics[width=\linewidth]{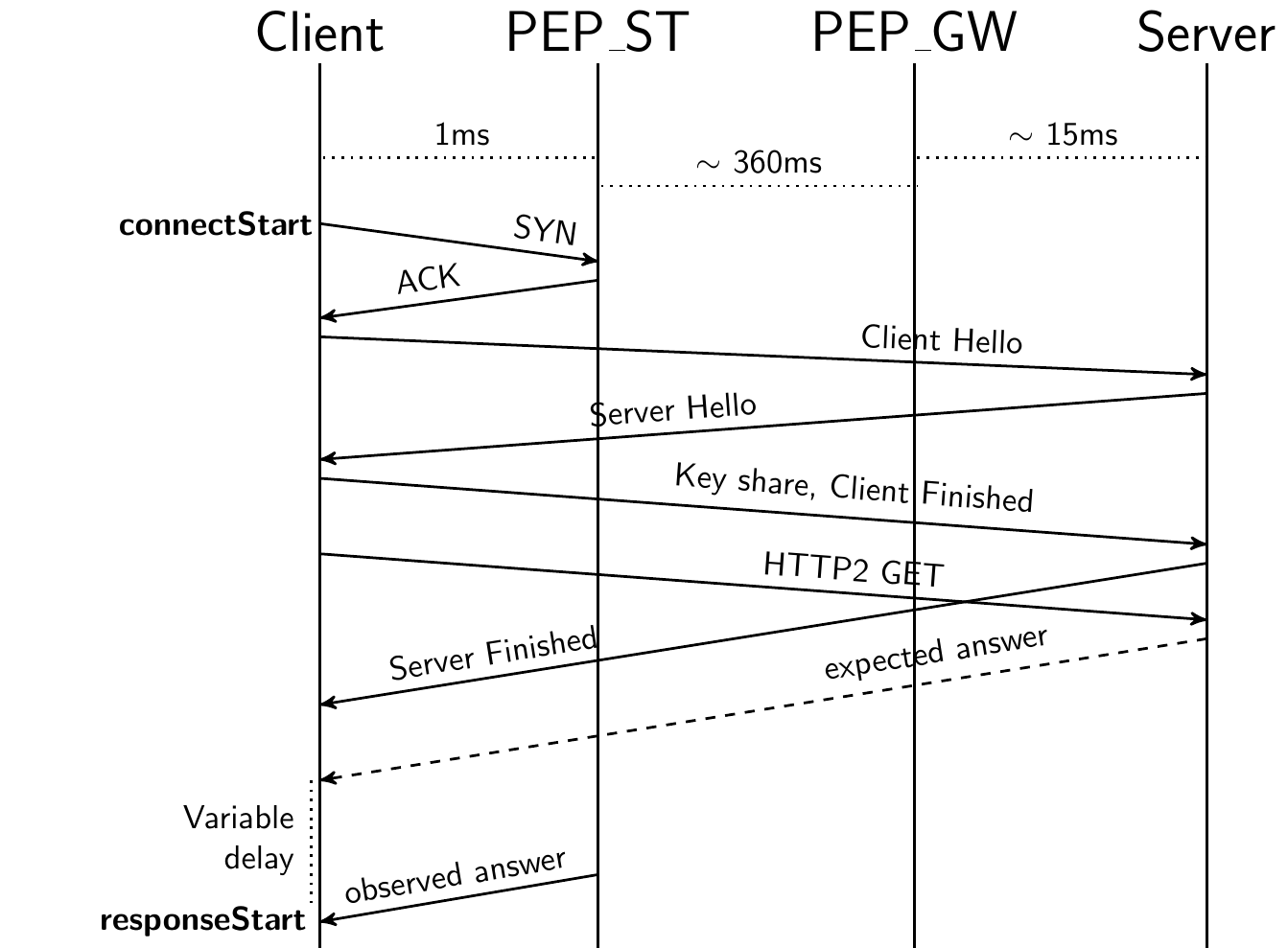}
	\caption{HTT handshake sequence. TCP handshake duration can be neglected thanks to PEP}
	\label{fig:results:handshake}
\end{figure}
\begin{figure}
	\centering
	\includegraphics[width=0.7\linewidth]{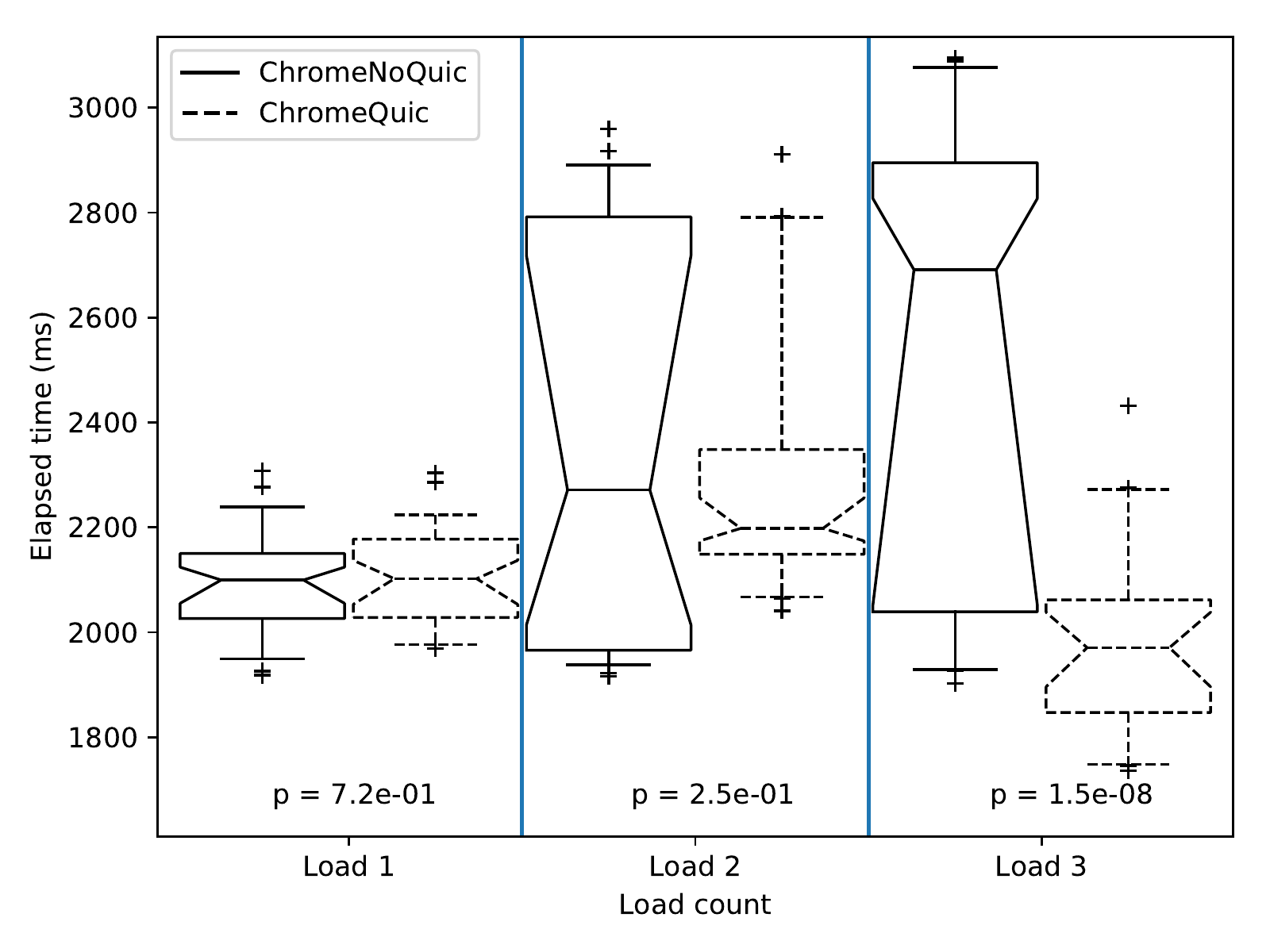}
	\caption{Time to \texttt{responseStart} (first HTTP byte) for target A on a SATCOM access}
	\label{fig:results:ttr}
\end{figure}
\begin{figure}
	\centering
	\includegraphics[width=0.7\linewidth]{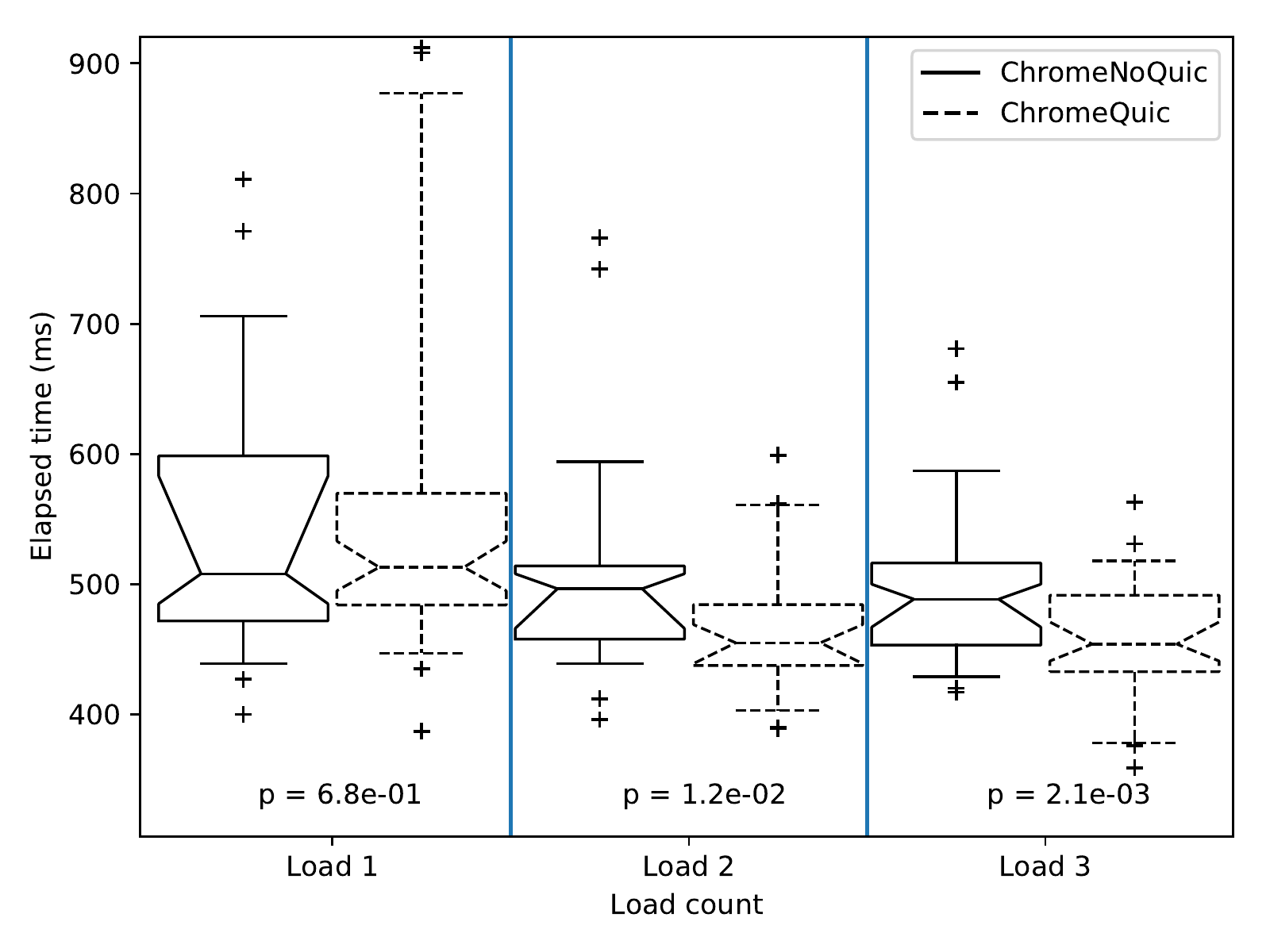}
	\caption{Time to \texttt{responseStart} (first HTTP byte) for target A on a 4G access}
	\label{fig:results:ttr_4g}
\end{figure}

We just saw that GQUIC encryption prevents any proxy to split the connection, which results in long CC \emph{Startup} phase. But GQUIC was also designed to reduce the handshake duration. In this section, we focus on that phase of the download: from the first packet sent by the client to the first HTTP response bytes received. Results are shown in Figure \ref{fig:results:ttr} that presents the notch boxes for the TTR metric. 


Loads 2 and 3 show a high dispersion for \texttt{ChromeNoQuic} and incompatible confidence intervals for \texttt{ChromeQuic}. It questions the hypothesis that learning mechanisms are performed within one load. 
On a 4G access network (Figure \ref{fig:results:ttr_4g}), this behavior cannot be seen : we note a gain between load 1 and the followings but loads 2 and 3 show low dispersion and compatible confidence intervals. 
We assume that the SATCOM ISP policy might disturb the results based on recent traffic history. 

One could expect that HQU would gain at least one RTT during the handshake, our results do not reflect that expectation. 
To understand why, we first  need to note that neither GQUIC nor TLS1.2 use connection resumption since the browser is restarted each time. However, analysis of traffic indicates that \texttt{Chrome} is using TLS1.2 False Start~\cite{tcp-false-start}. The observed HTT handshake is presented in Figure \ref{fig:results:handshake}. The TCP handshake is performed with the immediate PEP on the path. Its duration can be neglected compared to the rest of the sequence. The TLS1.2 Client Hello packet does not suffer any extra delay. So, it appears that the middle segment use a proprietary protocol or already existing connections and no handshake is required between the two PEP.  As a consequence we expect to receive the first byte with the HTT stack (\textit{i.e.} $TTR_{HTT}$) within twice the E2E RTT. In reality, we observe a high-dispersion extra-delay before receiving the HTTP answer (see Figure \ref{fig:results:handshake}). We put those results in relation with the same dispersive delay observed in~\cite{long-look, quic-satcom} and we could not identify with certainty its origin. That being said, the analysis can rely on the median and maximum values of the metric to drive trends.

For the HQU stack, since connection resumption is not used, GQUIC will perform 1 RTT handshakes (Figure 4 of~\cite{quic-deploy-doc}). Here again we can expect : $TTR_{HQU} \approx 2 RTT_{E2E}$. In conclusion, thanks to the distributed PEP, to TLS False Start and despite the 1 RTT handshake in GQUIC, both the HTT and the HQU stacks present the same theoretical TTR. In practice, HQU is here slightly faster. 

To further justify our analysis, we compare the TTR on a 4G access network (Figure \ref{fig:results:ttr_4g}).
In this network, the TCP handshake cannot be neglected anymore because PEP are not deployed. Thus, $TTR_{HTT} = 3 RTT_{E2E}$, \textit{i.e.}, HQU is faster by at least one E2E RTT which is consistent with Figure~\ref{fig:results:seq_numbers_4g} and Table~\ref{tab:qos} as the measured TTR gap is around $90$\,ms. It explains why, when compared to HTT, HQU might present better performances on a mobile network than on a SATCOM access.

\subsection{PLT for a small page}

\begin{figure}
	\centering
	\includegraphics[width=0.7\linewidth]{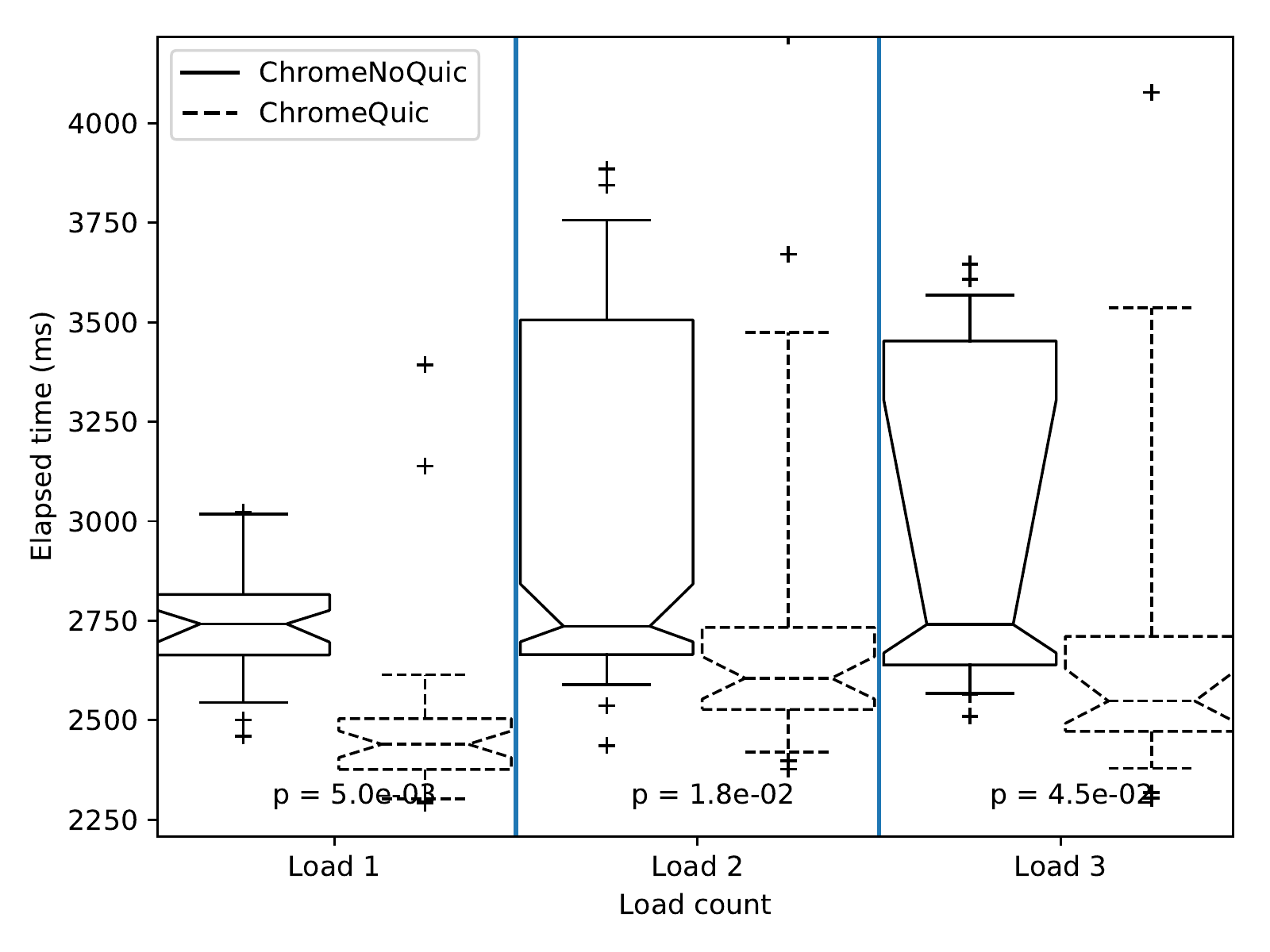}
	\caption{PLT for target B}
	\label{fig:results:plt:target_b}
\end{figure}

In this section, we aim at assessing the impact of the target size on the above mentioned conclusions and further analyze the impact of the ratio between the BDP, the IW. Figure \ref{fig:results:plt:target_b} presents those results.
 
Target B can be sent within an initial congestion window (See Table \ref{tab:odg}). \texttt{ChromeQuic} exhibits a better PLT for loads 2 and 3 of target B than \texttt{ChromeNoQuic}. Indeed, the CC does not need to probe the link and the main contributor of the PLT should be the TTR. Moreover, as we saw in section \ref{sec:results:handshake}, HQU generally fires the \texttt{responseStart} event slightly before HTT. 

That being said, as opposed to target A, objects of target B are located on \texttt{www.google.com}\footnote{Given by analysis of the internal log of \texttt{Chrome}}. These objects are downloaded by \texttt{ChromeQuic} using an already existing GQUIC connection. This is due to the initial connections to \texttt{www.google.com} that \texttt{Chrome} opens when starting. They enable \texttt{ChromeQuic} to discover GQUIC support and even to reuse the previously opened GQUIC connection to fetch the two objects of page B. \textit{I.e.}, by reducing the duration of several handshakes to several servers, the gains are summed up and the PLT is reduced. However, we note that the gain is limited compared to the PLT difference with target A. 

%% file: Figure_paper_PLT_A.pgf
\begingroup%
\makeatletter%
\begin{pgfpicture}%
\pgfpathrectangle{\pgfpointorigin}{\pgfqpoint{6.400000in}{4.800000in}}%
\pgfusepath{use as bounding box, clip}%
\begin{pgfscope}%
\pgfsetbuttcap%
\pgfsetmiterjoin%
\definecolor{currentfill}{rgb}{1.000000,1.000000,1.000000}%
\pgfsetfillcolor{currentfill}%
\pgfsetlinewidth{0.000000pt}%
\definecolor{currentstroke}{rgb}{1.000000,1.000000,1.000000}%
\pgfsetstrokecolor{currentstroke}%
\pgfsetdash{}{0pt}%
\pgfpathmoveto{\pgfqpoint{0.000000in}{0.000000in}}%
\pgfpathlineto{\pgfqpoint{6.400000in}{0.000000in}}%
\pgfpathlineto{\pgfqpoint{6.400000in}{4.800000in}}%
\pgfpathlineto{\pgfqpoint{0.000000in}{4.800000in}}%
\pgfpathclose%
\pgfusepath{fill}%
\end{pgfscope}%
\begin{pgfscope}%
\pgfsetbuttcap%
\pgfsetmiterjoin%
\definecolor{currentfill}{rgb}{1.000000,1.000000,1.000000}%
\pgfsetfillcolor{currentfill}%
\pgfsetlinewidth{0.000000pt}%
\definecolor{currentstroke}{rgb}{0.000000,0.000000,0.000000}%
\pgfsetstrokecolor{currentstroke}%
\pgfsetstrokeopacity{0.000000}%
\pgfsetdash{}{0pt}%
\pgfpathmoveto{\pgfqpoint{0.881528in}{0.582778in}}%
\pgfpathlineto{\pgfqpoint{6.215000in}{0.582778in}}%
\pgfpathlineto{\pgfqpoint{6.215000in}{4.615000in}}%
\pgfpathlineto{\pgfqpoint{0.881528in}{4.615000in}}%
\pgfpathclose%
\pgfusepath{fill}%
\end{pgfscope}%
\begin{pgfscope}%
\pgfsetbuttcap%
\pgfsetroundjoin%
\definecolor{currentfill}{rgb}{0.000000,0.000000,0.000000}%
\pgfsetfillcolor{currentfill}%
\pgfsetlinewidth{0.803000pt}%
\definecolor{currentstroke}{rgb}{0.000000,0.000000,0.000000}%
\pgfsetstrokecolor{currentstroke}%
\pgfsetdash{}{0pt}%
\pgfsys@defobject{currentmarker}{\pgfqpoint{0.000000in}{-0.048611in}}{\pgfqpoint{0.000000in}{0.000000in}}{%
\pgfpathmoveto{\pgfqpoint{0.000000in}{0.000000in}}%
\pgfpathlineto{\pgfqpoint{0.000000in}{-0.048611in}}%
\pgfusepath{stroke,fill}%
}%
\begin{pgfscope}%
\pgfsys@transformshift{1.771117in}{0.582778in}%
\pgfsys@useobject{currentmarker}{}%
\end{pgfscope}%
\end{pgfscope}%
\begin{pgfscope}%
\pgftext[x=1.771117in,y=0.485556in,,top]{\sffamily\fontsize{10.000000}{12.000000}\selectfont Load 1}%
\end{pgfscope}%
\begin{pgfscope}%
\pgfsetbuttcap%
\pgfsetroundjoin%
\definecolor{currentfill}{rgb}{0.000000,0.000000,0.000000}%
\pgfsetfillcolor{currentfill}%
\pgfsetlinewidth{0.803000pt}%
\definecolor{currentstroke}{rgb}{0.000000,0.000000,0.000000}%
\pgfsetstrokecolor{currentstroke}%
\pgfsetdash{}{0pt}%
\pgfsys@defobject{currentmarker}{\pgfqpoint{0.000000in}{-0.048611in}}{\pgfqpoint{0.000000in}{0.000000in}}{%
\pgfpathmoveto{\pgfqpoint{0.000000in}{0.000000in}}%
\pgfpathlineto{\pgfqpoint{0.000000in}{-0.048611in}}%
\pgfusepath{stroke,fill}%
}%
\begin{pgfscope}%
\pgfsys@transformshift{3.548355in}{0.582778in}%
\pgfsys@useobject{currentmarker}{}%
\end{pgfscope}%
\end{pgfscope}%
\begin{pgfscope}%
\pgftext[x=3.548355in,y=0.485556in,,top]{\sffamily\fontsize{10.000000}{12.000000}\selectfont Load 2}%
\end{pgfscope}%
\begin{pgfscope}%
\pgfsetbuttcap%
\pgfsetroundjoin%
\definecolor{currentfill}{rgb}{0.000000,0.000000,0.000000}%
\pgfsetfillcolor{currentfill}%
\pgfsetlinewidth{0.803000pt}%
\definecolor{currentstroke}{rgb}{0.000000,0.000000,0.000000}%
\pgfsetstrokecolor{currentstroke}%
\pgfsetdash{}{0pt}%
\pgfsys@defobject{currentmarker}{\pgfqpoint{0.000000in}{-0.048611in}}{\pgfqpoint{0.000000in}{0.000000in}}{%
\pgfpathmoveto{\pgfqpoint{0.000000in}{0.000000in}}%
\pgfpathlineto{\pgfqpoint{0.000000in}{-0.048611in}}%
\pgfusepath{stroke,fill}%
}%
\begin{pgfscope}%
\pgfsys@transformshift{5.325592in}{0.582778in}%
\pgfsys@useobject{currentmarker}{}%
\end{pgfscope}%
\end{pgfscope}%
\begin{pgfscope}%
\pgftext[x=5.325592in,y=0.485556in,,top]{\sffamily\fontsize{10.000000}{12.000000}\selectfont Load 3}%
\end{pgfscope}%
\begin{pgfscope}%
\pgftext[x=3.548264in,y=0.295587in,,top]{\sffamily\fontsize{10.000000}{12.000000}\selectfont Load count}%
\end{pgfscope}%
\begin{pgfscope}%
\pgfsetbuttcap%
\pgfsetroundjoin%
\definecolor{currentfill}{rgb}{0.000000,0.000000,0.000000}%
\pgfsetfillcolor{currentfill}%
\pgfsetlinewidth{0.803000pt}%
\definecolor{currentstroke}{rgb}{0.000000,0.000000,0.000000}%
\pgfsetstrokecolor{currentstroke}%
\pgfsetdash{}{0pt}%
\pgfsys@defobject{currentmarker}{\pgfqpoint{-0.048611in}{0.000000in}}{\pgfqpoint{0.000000in}{0.000000in}}{%
\pgfpathmoveto{\pgfqpoint{0.000000in}{0.000000in}}%
\pgfpathlineto{\pgfqpoint{-0.048611in}{0.000000in}}%
\pgfusepath{stroke,fill}%
}%
\begin{pgfscope}%
\pgfsys@transformshift{0.881528in}{0.994851in}%
\pgfsys@useobject{currentmarker}{}%
\end{pgfscope}%
\end{pgfscope}%
\begin{pgfscope}%
\pgftext[x=0.430844in,y=0.942089in,left,base]{\sffamily\fontsize{10.000000}{12.000000}\selectfont 4000}%
\end{pgfscope}%
\begin{pgfscope}%
\pgfsetbuttcap%
\pgfsetroundjoin%
\definecolor{currentfill}{rgb}{0.000000,0.000000,0.000000}%
\pgfsetfillcolor{currentfill}%
\pgfsetlinewidth{0.803000pt}%
\definecolor{currentstroke}{rgb}{0.000000,0.000000,0.000000}%
\pgfsetstrokecolor{currentstroke}%
\pgfsetdash{}{0pt}%
\pgfsys@defobject{currentmarker}{\pgfqpoint{-0.048611in}{0.000000in}}{\pgfqpoint{0.000000in}{0.000000in}}{%
\pgfpathmoveto{\pgfqpoint{0.000000in}{0.000000in}}%
\pgfpathlineto{\pgfqpoint{-0.048611in}{0.000000in}}%
\pgfusepath{stroke,fill}%
}%
\begin{pgfscope}%
\pgfsys@transformshift{0.881528in}{1.540192in}%
\pgfsys@useobject{currentmarker}{}%
\end{pgfscope}%
\end{pgfscope}%
\begin{pgfscope}%
\pgftext[x=0.430844in,y=1.487431in,left,base]{\sffamily\fontsize{10.000000}{12.000000}\selectfont 5000}%
\end{pgfscope}%
\begin{pgfscope}%
\pgfsetbuttcap%
\pgfsetroundjoin%
\definecolor{currentfill}{rgb}{0.000000,0.000000,0.000000}%
\pgfsetfillcolor{currentfill}%
\pgfsetlinewidth{0.803000pt}%
\definecolor{currentstroke}{rgb}{0.000000,0.000000,0.000000}%
\pgfsetstrokecolor{currentstroke}%
\pgfsetdash{}{0pt}%
\pgfsys@defobject{currentmarker}{\pgfqpoint{-0.048611in}{0.000000in}}{\pgfqpoint{0.000000in}{0.000000in}}{%
\pgfpathmoveto{\pgfqpoint{0.000000in}{0.000000in}}%
\pgfpathlineto{\pgfqpoint{-0.048611in}{0.000000in}}%
\pgfusepath{stroke,fill}%
}%
\begin{pgfscope}%
\pgfsys@transformshift{0.881528in}{2.085533in}%
\pgfsys@useobject{currentmarker}{}%
\end{pgfscope}%
\end{pgfscope}%
\begin{pgfscope}%
\pgftext[x=0.430844in,y=2.032772in,left,base]{\sffamily\fontsize{10.000000}{12.000000}\selectfont 6000}%
\end{pgfscope}%
\begin{pgfscope}%
\pgfsetbuttcap%
\pgfsetroundjoin%
\definecolor{currentfill}{rgb}{0.000000,0.000000,0.000000}%
\pgfsetfillcolor{currentfill}%
\pgfsetlinewidth{0.803000pt}%
\definecolor{currentstroke}{rgb}{0.000000,0.000000,0.000000}%
\pgfsetstrokecolor{currentstroke}%
\pgfsetdash{}{0pt}%
\pgfsys@defobject{currentmarker}{\pgfqpoint{-0.048611in}{0.000000in}}{\pgfqpoint{0.000000in}{0.000000in}}{%
\pgfpathmoveto{\pgfqpoint{0.000000in}{0.000000in}}%
\pgfpathlineto{\pgfqpoint{-0.048611in}{0.000000in}}%
\pgfusepath{stroke,fill}%
}%
\begin{pgfscope}%
\pgfsys@transformshift{0.881528in}{2.630875in}%
\pgfsys@useobject{currentmarker}{}%
\end{pgfscope}%
\end{pgfscope}%
\begin{pgfscope}%
\pgftext[x=0.430844in,y=2.578113in,left,base]{\sffamily\fontsize{10.000000}{12.000000}\selectfont 7000}%
\end{pgfscope}%
\begin{pgfscope}%
\pgfsetbuttcap%
\pgfsetroundjoin%
\definecolor{currentfill}{rgb}{0.000000,0.000000,0.000000}%
\pgfsetfillcolor{currentfill}%
\pgfsetlinewidth{0.803000pt}%
\definecolor{currentstroke}{rgb}{0.000000,0.000000,0.000000}%
\pgfsetstrokecolor{currentstroke}%
\pgfsetdash{}{0pt}%
\pgfsys@defobject{currentmarker}{\pgfqpoint{-0.048611in}{0.000000in}}{\pgfqpoint{0.000000in}{0.000000in}}{%
\pgfpathmoveto{\pgfqpoint{0.000000in}{0.000000in}}%
\pgfpathlineto{\pgfqpoint{-0.048611in}{0.000000in}}%
\pgfusepath{stroke,fill}%
}%
\begin{pgfscope}%
\pgfsys@transformshift{0.881528in}{3.176216in}%
\pgfsys@useobject{currentmarker}{}%
\end{pgfscope}%
\end{pgfscope}%
\begin{pgfscope}%
\pgftext[x=0.430844in,y=3.123454in,left,base]{\sffamily\fontsize{10.000000}{12.000000}\selectfont 8000}%
\end{pgfscope}%
\begin{pgfscope}%
\pgfsetbuttcap%
\pgfsetroundjoin%
\definecolor{currentfill}{rgb}{0.000000,0.000000,0.000000}%
\pgfsetfillcolor{currentfill}%
\pgfsetlinewidth{0.803000pt}%
\definecolor{currentstroke}{rgb}{0.000000,0.000000,0.000000}%
\pgfsetstrokecolor{currentstroke}%
\pgfsetdash{}{0pt}%
\pgfsys@defobject{currentmarker}{\pgfqpoint{-0.048611in}{0.000000in}}{\pgfqpoint{0.000000in}{0.000000in}}{%
\pgfpathmoveto{\pgfqpoint{0.000000in}{0.000000in}}%
\pgfpathlineto{\pgfqpoint{-0.048611in}{0.000000in}}%
\pgfusepath{stroke,fill}%
}%
\begin{pgfscope}%
\pgfsys@transformshift{0.881528in}{3.721557in}%
\pgfsys@useobject{currentmarker}{}%
\end{pgfscope}%
\end{pgfscope}%
\begin{pgfscope}%
\pgftext[x=0.430844in,y=3.668796in,left,base]{\sffamily\fontsize{10.000000}{12.000000}\selectfont 9000}%
\end{pgfscope}%
\begin{pgfscope}%
\pgfsetbuttcap%
\pgfsetroundjoin%
\definecolor{currentfill}{rgb}{0.000000,0.000000,0.000000}%
\pgfsetfillcolor{currentfill}%
\pgfsetlinewidth{0.803000pt}%
\definecolor{currentstroke}{rgb}{0.000000,0.000000,0.000000}%
\pgfsetstrokecolor{currentstroke}%
\pgfsetdash{}{0pt}%
\pgfsys@defobject{currentmarker}{\pgfqpoint{-0.048611in}{0.000000in}}{\pgfqpoint{0.000000in}{0.000000in}}{%
\pgfpathmoveto{\pgfqpoint{0.000000in}{0.000000in}}%
\pgfpathlineto{\pgfqpoint{-0.048611in}{0.000000in}}%
\pgfusepath{stroke,fill}%
}%
\begin{pgfscope}%
\pgfsys@transformshift{0.881528in}{4.266898in}%
\pgfsys@useobject{currentmarker}{}%
\end{pgfscope}%
\end{pgfscope}%
\begin{pgfscope}%
\pgftext[x=0.342479in,y=4.214137in,left,base]{\sffamily\fontsize{10.000000}{12.000000}\selectfont 10000}%
\end{pgfscope}%
\begin{pgfscope}%
\pgftext[x=0.286923in,y=2.598889in,,bottom,rotate=90.000000]{\sffamily\fontsize{10.000000}{12.000000}\selectfont Elapsed time (ms)}%
\end{pgfscope}%
\begin{pgfscope}%
\pgfpathrectangle{\pgfqpoint{0.881528in}{0.582778in}}{\pgfqpoint{5.333472in}{4.032222in}}%
\pgfusepath{clip}%
\pgfsetrectcap%
\pgfsetroundjoin%
\pgfsetlinewidth{1.003750pt}%
\definecolor{currentstroke}{rgb}{0.000000,0.000000,0.000000}%
\pgfsetstrokecolor{currentstroke}%
\pgfsetdash{}{0pt}%
\pgfpathmoveto{\pgfqpoint{0.904714in}{1.001940in}}%
\pgfpathlineto{\pgfqpoint{1.748902in}{1.001940in}}%
\pgfpathlineto{\pgfqpoint{1.748902in}{1.022118in}}%
\pgfpathlineto{\pgfqpoint{1.537855in}{1.042295in}}%
\pgfpathlineto{\pgfqpoint{1.748902in}{1.078833in}}%
\pgfpathlineto{\pgfqpoint{1.748902in}{1.086604in}}%
\pgfpathlineto{\pgfqpoint{0.904714in}{1.086604in}}%
\pgfpathlineto{\pgfqpoint{0.904714in}{1.078833in}}%
\pgfpathlineto{\pgfqpoint{1.115761in}{1.042295in}}%
\pgfpathlineto{\pgfqpoint{0.904714in}{1.022118in}}%
\pgfpathlineto{\pgfqpoint{0.904714in}{1.001940in}}%
\pgfusepath{stroke}%
\end{pgfscope}%
\begin{pgfscope}%
\pgfpathrectangle{\pgfqpoint{0.881528in}{0.582778in}}{\pgfqpoint{5.333472in}{4.032222in}}%
\pgfusepath{clip}%
\pgfsetrectcap%
\pgfsetroundjoin%
\pgfsetlinewidth{1.003750pt}%
\definecolor{currentstroke}{rgb}{0.000000,0.000000,0.000000}%
\pgfsetstrokecolor{currentstroke}%
\pgfsetdash{}{0pt}%
\pgfpathmoveto{\pgfqpoint{1.326808in}{1.001940in}}%
\pgfpathlineto{\pgfqpoint{1.326808in}{0.975219in}}%
\pgfusepath{stroke}%
\end{pgfscope}%
\begin{pgfscope}%
\pgfpathrectangle{\pgfqpoint{0.881528in}{0.582778in}}{\pgfqpoint{5.333472in}{4.032222in}}%
\pgfusepath{clip}%
\pgfsetrectcap%
\pgfsetroundjoin%
\pgfsetlinewidth{1.003750pt}%
\definecolor{currentstroke}{rgb}{0.000000,0.000000,0.000000}%
\pgfsetstrokecolor{currentstroke}%
\pgfsetdash{}{0pt}%
\pgfpathmoveto{\pgfqpoint{1.326808in}{1.086604in}}%
\pgfpathlineto{\pgfqpoint{1.326808in}{1.172632in}}%
\pgfusepath{stroke}%
\end{pgfscope}%
\begin{pgfscope}%
\pgfpathrectangle{\pgfqpoint{0.881528in}{0.582778in}}{\pgfqpoint{5.333472in}{4.032222in}}%
\pgfusepath{clip}%
\pgfsetrectcap%
\pgfsetroundjoin%
\pgfsetlinewidth{1.003750pt}%
\definecolor{currentstroke}{rgb}{0.000000,0.000000,0.000000}%
\pgfsetstrokecolor{currentstroke}%
\pgfsetdash{}{0pt}%
\pgfpathmoveto{\pgfqpoint{1.115761in}{0.975219in}}%
\pgfpathlineto{\pgfqpoint{1.537855in}{0.975219in}}%
\pgfusepath{stroke}%
\end{pgfscope}%
\begin{pgfscope}%
\pgfpathrectangle{\pgfqpoint{0.881528in}{0.582778in}}{\pgfqpoint{5.333472in}{4.032222in}}%
\pgfusepath{clip}%
\pgfsetrectcap%
\pgfsetroundjoin%
\pgfsetlinewidth{1.003750pt}%
\definecolor{currentstroke}{rgb}{0.000000,0.000000,0.000000}%
\pgfsetstrokecolor{currentstroke}%
\pgfsetdash{}{0pt}%
\pgfpathmoveto{\pgfqpoint{1.115761in}{1.172632in}}%
\pgfpathlineto{\pgfqpoint{1.537855in}{1.172632in}}%
\pgfusepath{stroke}%
\end{pgfscope}%
\begin{pgfscope}%
\pgfpathrectangle{\pgfqpoint{0.881528in}{0.582778in}}{\pgfqpoint{5.333472in}{4.032222in}}%
\pgfusepath{clip}%
\pgfsetbuttcap%
\pgfsetroundjoin%
\definecolor{currentfill}{rgb}{0.000000,0.000000,0.000000}%
\pgfsetfillcolor{currentfill}%
\pgfsetfillopacity{0.000000}%
\pgfsetlinewidth{1.003750pt}%
\definecolor{currentstroke}{rgb}{0.000000,0.000000,0.000000}%
\pgfsetstrokecolor{currentstroke}%
\pgfsetdash{}{0pt}%
\pgfsys@defobject{currentmarker}{\pgfqpoint{-0.041667in}{-0.041667in}}{\pgfqpoint{0.041667in}{0.041667in}}{%
\pgfpathmoveto{\pgfqpoint{-0.041667in}{0.000000in}}%
\pgfpathlineto{\pgfqpoint{0.041667in}{0.000000in}}%
\pgfpathmoveto{\pgfqpoint{0.000000in}{-0.041667in}}%
\pgfpathlineto{\pgfqpoint{0.000000in}{0.041667in}}%
\pgfusepath{stroke,fill}%
}%
\begin{pgfscope}%
\pgfsys@transformshift{1.326808in}{0.956677in}%
\pgfsys@useobject{currentmarker}{}%
\end{pgfscope}%
\begin{pgfscope}%
\pgfsys@transformshift{1.326808in}{0.935409in}%
\pgfsys@useobject{currentmarker}{}%
\end{pgfscope}%
\begin{pgfscope}%
\pgfsys@transformshift{1.326808in}{1.194991in}%
\pgfsys@useobject{currentmarker}{}%
\end{pgfscope}%
\begin{pgfscope}%
\pgfsys@transformshift{1.326808in}{1.206443in}%
\pgfsys@useobject{currentmarker}{}%
\end{pgfscope}%
\end{pgfscope}%
\begin{pgfscope}%
\pgfpathrectangle{\pgfqpoint{0.881528in}{0.582778in}}{\pgfqpoint{5.333472in}{4.032222in}}%
\pgfusepath{clip}%
\pgfsetrectcap%
\pgfsetroundjoin%
\pgfsetlinewidth{1.003750pt}%
\definecolor{currentstroke}{rgb}{0.000000,0.000000,0.000000}%
\pgfsetstrokecolor{currentstroke}%
\pgfsetdash{}{0pt}%
\pgfpathmoveto{\pgfqpoint{2.681952in}{0.983126in}}%
\pgfpathlineto{\pgfqpoint{3.526139in}{0.983126in}}%
\pgfpathlineto{\pgfqpoint{3.526139in}{1.037933in}}%
\pgfpathlineto{\pgfqpoint{3.315092in}{1.223349in}}%
\pgfpathlineto{\pgfqpoint{3.526139in}{1.413128in}}%
\pgfpathlineto{\pgfqpoint{3.526139in}{1.429351in}}%
\pgfpathlineto{\pgfqpoint{2.681952in}{1.429351in}}%
\pgfpathlineto{\pgfqpoint{2.681952in}{1.413128in}}%
\pgfpathlineto{\pgfqpoint{2.892998in}{1.223349in}}%
\pgfpathlineto{\pgfqpoint{2.681952in}{1.037933in}}%
\pgfpathlineto{\pgfqpoint{2.681952in}{0.983126in}}%
\pgfusepath{stroke}%
\end{pgfscope}%
\begin{pgfscope}%
\pgfpathrectangle{\pgfqpoint{0.881528in}{0.582778in}}{\pgfqpoint{5.333472in}{4.032222in}}%
\pgfusepath{clip}%
\pgfsetrectcap%
\pgfsetroundjoin%
\pgfsetlinewidth{1.003750pt}%
\definecolor{currentstroke}{rgb}{0.000000,0.000000,0.000000}%
\pgfsetstrokecolor{currentstroke}%
\pgfsetdash{}{0pt}%
\pgfpathmoveto{\pgfqpoint{3.104045in}{0.983126in}}%
\pgfpathlineto{\pgfqpoint{3.104045in}{0.936499in}}%
\pgfusepath{stroke}%
\end{pgfscope}%
\begin{pgfscope}%
\pgfpathrectangle{\pgfqpoint{0.881528in}{0.582778in}}{\pgfqpoint{5.333472in}{4.032222in}}%
\pgfusepath{clip}%
\pgfsetrectcap%
\pgfsetroundjoin%
\pgfsetlinewidth{1.003750pt}%
\definecolor{currentstroke}{rgb}{0.000000,0.000000,0.000000}%
\pgfsetstrokecolor{currentstroke}%
\pgfsetdash{}{0pt}%
\pgfpathmoveto{\pgfqpoint{3.104045in}{1.429351in}}%
\pgfpathlineto{\pgfqpoint{3.104045in}{1.514016in}}%
\pgfusepath{stroke}%
\end{pgfscope}%
\begin{pgfscope}%
\pgfpathrectangle{\pgfqpoint{0.881528in}{0.582778in}}{\pgfqpoint{5.333472in}{4.032222in}}%
\pgfusepath{clip}%
\pgfsetrectcap%
\pgfsetroundjoin%
\pgfsetlinewidth{1.003750pt}%
\definecolor{currentstroke}{rgb}{0.000000,0.000000,0.000000}%
\pgfsetstrokecolor{currentstroke}%
\pgfsetdash{}{0pt}%
\pgfpathmoveto{\pgfqpoint{2.892998in}{0.936499in}}%
\pgfpathlineto{\pgfqpoint{3.315092in}{0.936499in}}%
\pgfusepath{stroke}%
\end{pgfscope}%
\begin{pgfscope}%
\pgfpathrectangle{\pgfqpoint{0.881528in}{0.582778in}}{\pgfqpoint{5.333472in}{4.032222in}}%
\pgfusepath{clip}%
\pgfsetrectcap%
\pgfsetroundjoin%
\pgfsetlinewidth{1.003750pt}%
\definecolor{currentstroke}{rgb}{0.000000,0.000000,0.000000}%
\pgfsetstrokecolor{currentstroke}%
\pgfsetdash{}{0pt}%
\pgfpathmoveto{\pgfqpoint{2.892998in}{1.514016in}}%
\pgfpathlineto{\pgfqpoint{3.315092in}{1.514016in}}%
\pgfusepath{stroke}%
\end{pgfscope}%
\begin{pgfscope}%
\pgfpathrectangle{\pgfqpoint{0.881528in}{0.582778in}}{\pgfqpoint{5.333472in}{4.032222in}}%
\pgfusepath{clip}%
\pgfsetbuttcap%
\pgfsetroundjoin%
\definecolor{currentfill}{rgb}{0.000000,0.000000,0.000000}%
\pgfsetfillcolor{currentfill}%
\pgfsetfillopacity{0.000000}%
\pgfsetlinewidth{1.003750pt}%
\definecolor{currentstroke}{rgb}{0.000000,0.000000,0.000000}%
\pgfsetstrokecolor{currentstroke}%
\pgfsetdash{}{0pt}%
\pgfsys@defobject{currentmarker}{\pgfqpoint{-0.041667in}{-0.041667in}}{\pgfqpoint{0.041667in}{0.041667in}}{%
\pgfpathmoveto{\pgfqpoint{-0.041667in}{0.000000in}}%
\pgfpathlineto{\pgfqpoint{0.041667in}{0.000000in}}%
\pgfpathmoveto{\pgfqpoint{0.000000in}{-0.041667in}}%
\pgfpathlineto{\pgfqpoint{0.000000in}{0.041667in}}%
\pgfusepath{stroke,fill}%
}%
\begin{pgfscope}%
\pgfsys@transformshift{3.104045in}{0.932137in}%
\pgfsys@useobject{currentmarker}{}%
\end{pgfscope}%
\begin{pgfscope}%
\pgfsys@transformshift{3.104045in}{1.542919in}%
\pgfsys@useobject{currentmarker}{}%
\end{pgfscope}%
\begin{pgfscope}%
\pgfsys@transformshift{3.104045in}{1.619267in}%
\pgfsys@useobject{currentmarker}{}%
\end{pgfscope}%
\end{pgfscope}%
\begin{pgfscope}%
\pgfpathrectangle{\pgfqpoint{0.881528in}{0.582778in}}{\pgfqpoint{5.333472in}{4.032222in}}%
\pgfusepath{clip}%
\pgfsetrectcap%
\pgfsetroundjoin%
\pgfsetlinewidth{1.003750pt}%
\definecolor{currentstroke}{rgb}{0.000000,0.000000,0.000000}%
\pgfsetstrokecolor{currentstroke}%
\pgfsetdash{}{0pt}%
\pgfpathmoveto{\pgfqpoint{4.459189in}{1.023890in}}%
\pgfpathlineto{\pgfqpoint{5.303376in}{1.023890in}}%
\pgfpathlineto{\pgfqpoint{5.303376in}{1.059201in}}%
\pgfpathlineto{\pgfqpoint{5.092329in}{1.385588in}}%
\pgfpathlineto{\pgfqpoint{5.303376in}{1.506926in}}%
\pgfpathlineto{\pgfqpoint{5.303376in}{1.534193in}}%
\pgfpathlineto{\pgfqpoint{4.459189in}{1.534193in}}%
\pgfpathlineto{\pgfqpoint{4.459189in}{1.506926in}}%
\pgfpathlineto{\pgfqpoint{4.670236in}{1.385588in}}%
\pgfpathlineto{\pgfqpoint{4.459189in}{1.059201in}}%
\pgfpathlineto{\pgfqpoint{4.459189in}{1.023890in}}%
\pgfusepath{stroke}%
\end{pgfscope}%
\begin{pgfscope}%
\pgfpathrectangle{\pgfqpoint{0.881528in}{0.582778in}}{\pgfqpoint{5.333472in}{4.032222in}}%
\pgfusepath{clip}%
\pgfsetrectcap%
\pgfsetroundjoin%
\pgfsetlinewidth{1.003750pt}%
\definecolor{currentstroke}{rgb}{0.000000,0.000000,0.000000}%
\pgfsetstrokecolor{currentstroke}%
\pgfsetdash{}{0pt}%
\pgfpathmoveto{\pgfqpoint{4.881282in}{1.023890in}}%
\pgfpathlineto{\pgfqpoint{4.881282in}{0.943043in}}%
\pgfusepath{stroke}%
\end{pgfscope}%
\begin{pgfscope}%
\pgfpathrectangle{\pgfqpoint{0.881528in}{0.582778in}}{\pgfqpoint{5.333472in}{4.032222in}}%
\pgfusepath{clip}%
\pgfsetrectcap%
\pgfsetroundjoin%
\pgfsetlinewidth{1.003750pt}%
\definecolor{currentstroke}{rgb}{0.000000,0.000000,0.000000}%
\pgfsetstrokecolor{currentstroke}%
\pgfsetdash{}{0pt}%
\pgfpathmoveto{\pgfqpoint{4.881282in}{1.534193in}}%
\pgfpathlineto{\pgfqpoint{4.881282in}{1.630173in}}%
\pgfusepath{stroke}%
\end{pgfscope}%
\begin{pgfscope}%
\pgfpathrectangle{\pgfqpoint{0.881528in}{0.582778in}}{\pgfqpoint{5.333472in}{4.032222in}}%
\pgfusepath{clip}%
\pgfsetrectcap%
\pgfsetroundjoin%
\pgfsetlinewidth{1.003750pt}%
\definecolor{currentstroke}{rgb}{0.000000,0.000000,0.000000}%
\pgfsetstrokecolor{currentstroke}%
\pgfsetdash{}{0pt}%
\pgfpathmoveto{\pgfqpoint{4.670236in}{0.943043in}}%
\pgfpathlineto{\pgfqpoint{5.092329in}{0.943043in}}%
\pgfusepath{stroke}%
\end{pgfscope}%
\begin{pgfscope}%
\pgfpathrectangle{\pgfqpoint{0.881528in}{0.582778in}}{\pgfqpoint{5.333472in}{4.032222in}}%
\pgfusepath{clip}%
\pgfsetrectcap%
\pgfsetroundjoin%
\pgfsetlinewidth{1.003750pt}%
\definecolor{currentstroke}{rgb}{0.000000,0.000000,0.000000}%
\pgfsetstrokecolor{currentstroke}%
\pgfsetdash{}{0pt}%
\pgfpathmoveto{\pgfqpoint{4.670236in}{1.630173in}}%
\pgfpathlineto{\pgfqpoint{5.092329in}{1.630173in}}%
\pgfusepath{stroke}%
\end{pgfscope}%
\begin{pgfscope}%
\pgfpathrectangle{\pgfqpoint{0.881528in}{0.582778in}}{\pgfqpoint{5.333472in}{4.032222in}}%
\pgfusepath{clip}%
\pgfsetbuttcap%
\pgfsetroundjoin%
\definecolor{currentfill}{rgb}{0.000000,0.000000,0.000000}%
\pgfsetfillcolor{currentfill}%
\pgfsetfillopacity{0.000000}%
\pgfsetlinewidth{1.003750pt}%
\definecolor{currentstroke}{rgb}{0.000000,0.000000,0.000000}%
\pgfsetstrokecolor{currentstroke}%
\pgfsetdash{}{0pt}%
\pgfsys@defobject{currentmarker}{\pgfqpoint{-0.041667in}{-0.041667in}}{\pgfqpoint{0.041667in}{0.041667in}}{%
\pgfpathmoveto{\pgfqpoint{-0.041667in}{0.000000in}}%
\pgfpathlineto{\pgfqpoint{0.041667in}{0.000000in}}%
\pgfpathmoveto{\pgfqpoint{0.000000in}{-0.041667in}}%
\pgfpathlineto{\pgfqpoint{0.000000in}{0.041667in}}%
\pgfusepath{stroke,fill}%
}%
\begin{pgfscope}%
\pgfsys@transformshift{4.881282in}{0.922320in}%
\pgfsys@useobject{currentmarker}{}%
\end{pgfscope}%
\begin{pgfscope}%
\pgfsys@transformshift{4.881282in}{0.936499in}%
\pgfsys@useobject{currentmarker}{}%
\end{pgfscope}%
\begin{pgfscope}%
\pgfsys@transformshift{4.881282in}{1.902299in}%
\pgfsys@useobject{currentmarker}{}%
\end{pgfscope}%
\begin{pgfscope}%
\pgfsys@transformshift{4.881282in}{1.673255in}%
\pgfsys@useobject{currentmarker}{}%
\end{pgfscope}%
\end{pgfscope}%
\begin{pgfscope}%
\pgfpathrectangle{\pgfqpoint{0.881528in}{0.582778in}}{\pgfqpoint{5.333472in}{4.032222in}}%
\pgfusepath{clip}%
\pgfsetbuttcap%
\pgfsetroundjoin%
\pgfsetlinewidth{1.003750pt}%
\definecolor{currentstroke}{rgb}{0.000000,0.000000,0.000000}%
\pgfsetstrokecolor{currentstroke}%
\pgfsetdash{{3.700000pt}{1.600000pt}}{0.000000pt}%
\pgfpathmoveto{\pgfqpoint{1.793333in}{1.002213in}}%
\pgfpathlineto{\pgfqpoint{2.637521in}{1.002213in}}%
\pgfpathlineto{\pgfqpoint{2.637521in}{1.020482in}}%
\pgfpathlineto{\pgfqpoint{2.426474in}{1.056747in}}%
\pgfpathlineto{\pgfqpoint{2.637521in}{1.084287in}}%
\pgfpathlineto{\pgfqpoint{2.637521in}{1.101874in}}%
\pgfpathlineto{\pgfqpoint{1.793333in}{1.101874in}}%
\pgfpathlineto{\pgfqpoint{1.793333in}{1.084287in}}%
\pgfpathlineto{\pgfqpoint{2.004380in}{1.056747in}}%
\pgfpathlineto{\pgfqpoint{1.793333in}{1.020482in}}%
\pgfpathlineto{\pgfqpoint{1.793333in}{1.002213in}}%
\pgfusepath{stroke}%
\end{pgfscope}%
\begin{pgfscope}%
\pgfpathrectangle{\pgfqpoint{0.881528in}{0.582778in}}{\pgfqpoint{5.333472in}{4.032222in}}%
\pgfusepath{clip}%
\pgfsetbuttcap%
\pgfsetroundjoin%
\pgfsetlinewidth{1.003750pt}%
\definecolor{currentstroke}{rgb}{0.000000,0.000000,0.000000}%
\pgfsetstrokecolor{currentstroke}%
\pgfsetdash{{3.700000pt}{1.600000pt}}{0.000000pt}%
\pgfpathmoveto{\pgfqpoint{2.215427in}{1.002213in}}%
\pgfpathlineto{\pgfqpoint{2.215427in}{0.970856in}}%
\pgfusepath{stroke}%
\end{pgfscope}%
\begin{pgfscope}%
\pgfpathrectangle{\pgfqpoint{0.881528in}{0.582778in}}{\pgfqpoint{5.333472in}{4.032222in}}%
\pgfusepath{clip}%
\pgfsetbuttcap%
\pgfsetroundjoin%
\pgfsetlinewidth{1.003750pt}%
\definecolor{currentstroke}{rgb}{0.000000,0.000000,0.000000}%
\pgfsetstrokecolor{currentstroke}%
\pgfsetdash{{3.700000pt}{1.600000pt}}{0.000000pt}%
\pgfpathmoveto{\pgfqpoint{2.215427in}{1.101874in}}%
\pgfpathlineto{\pgfqpoint{2.215427in}{1.163907in}}%
\pgfusepath{stroke}%
\end{pgfscope}%
\begin{pgfscope}%
\pgfpathrectangle{\pgfqpoint{0.881528in}{0.582778in}}{\pgfqpoint{5.333472in}{4.032222in}}%
\pgfusepath{clip}%
\pgfsetbuttcap%
\pgfsetroundjoin%
\pgfsetlinewidth{1.003750pt}%
\definecolor{currentstroke}{rgb}{0.000000,0.000000,0.000000}%
\pgfsetstrokecolor{currentstroke}%
\pgfsetdash{{3.700000pt}{1.600000pt}}{0.000000pt}%
\pgfpathmoveto{\pgfqpoint{2.004380in}{0.970856in}}%
\pgfpathlineto{\pgfqpoint{2.426474in}{0.970856in}}%
\pgfusepath{stroke}%
\end{pgfscope}%
\begin{pgfscope}%
\pgfpathrectangle{\pgfqpoint{0.881528in}{0.582778in}}{\pgfqpoint{5.333472in}{4.032222in}}%
\pgfusepath{clip}%
\pgfsetbuttcap%
\pgfsetroundjoin%
\pgfsetlinewidth{1.003750pt}%
\definecolor{currentstroke}{rgb}{0.000000,0.000000,0.000000}%
\pgfsetstrokecolor{currentstroke}%
\pgfsetdash{{3.700000pt}{1.600000pt}}{0.000000pt}%
\pgfpathmoveto{\pgfqpoint{2.004380in}{1.163907in}}%
\pgfpathlineto{\pgfqpoint{2.426474in}{1.163907in}}%
\pgfusepath{stroke}%
\end{pgfscope}%
\begin{pgfscope}%
\pgfpathrectangle{\pgfqpoint{0.881528in}{0.582778in}}{\pgfqpoint{5.333472in}{4.032222in}}%
\pgfusepath{clip}%
\pgfsetbuttcap%
\pgfsetroundjoin%
\definecolor{currentfill}{rgb}{0.000000,0.000000,0.000000}%
\pgfsetfillcolor{currentfill}%
\pgfsetfillopacity{0.000000}%
\pgfsetlinewidth{1.003750pt}%
\definecolor{currentstroke}{rgb}{0.000000,0.000000,0.000000}%
\pgfsetstrokecolor{currentstroke}%
\pgfsetdash{}{0pt}%
\pgfsys@defobject{currentmarker}{\pgfqpoint{-0.041667in}{-0.041667in}}{\pgfqpoint{0.041667in}{0.041667in}}{%
\pgfpathmoveto{\pgfqpoint{-0.041667in}{0.000000in}}%
\pgfpathlineto{\pgfqpoint{0.041667in}{0.000000in}}%
\pgfpathmoveto{\pgfqpoint{0.000000in}{-0.041667in}}%
\pgfpathlineto{\pgfqpoint{0.000000in}{0.041667in}}%
\pgfusepath{stroke,fill}%
}%
\begin{pgfscope}%
\pgfsys@transformshift{2.215427in}{0.966493in}%
\pgfsys@useobject{currentmarker}{}%
\end{pgfscope}%
\begin{pgfscope}%
\pgfsys@transformshift{2.215427in}{0.961585in}%
\pgfsys@useobject{currentmarker}{}%
\end{pgfscope}%
\begin{pgfscope}%
\pgfsys@transformshift{2.215427in}{1.439849in}%
\pgfsys@useobject{currentmarker}{}%
\end{pgfscope}%
\begin{pgfscope}%
\pgfsys@transformshift{2.215427in}{1.271339in}%
\pgfsys@useobject{currentmarker}{}%
\end{pgfscope}%
\end{pgfscope}%
\begin{pgfscope}%
\pgfpathrectangle{\pgfqpoint{0.881528in}{0.582778in}}{\pgfqpoint{5.333472in}{4.032222in}}%
\pgfusepath{clip}%
\pgfsetbuttcap%
\pgfsetroundjoin%
\pgfsetlinewidth{1.003750pt}%
\definecolor{currentstroke}{rgb}{0.000000,0.000000,0.000000}%
\pgfsetstrokecolor{currentstroke}%
\pgfsetdash{{3.700000pt}{1.600000pt}}{0.000000pt}%
\pgfpathmoveto{\pgfqpoint{3.570570in}{3.564090in}}%
\pgfpathlineto{\pgfqpoint{4.414758in}{3.564090in}}%
\pgfpathlineto{\pgfqpoint{4.414758in}{3.601037in}}%
\pgfpathlineto{\pgfqpoint{4.203711in}{3.659934in}}%
\pgfpathlineto{\pgfqpoint{4.414758in}{3.722648in}}%
\pgfpathlineto{\pgfqpoint{4.414758in}{3.764639in}}%
\pgfpathlineto{\pgfqpoint{3.570570in}{3.764639in}}%
\pgfpathlineto{\pgfqpoint{3.570570in}{3.722648in}}%
\pgfpathlineto{\pgfqpoint{3.781617in}{3.659934in}}%
\pgfpathlineto{\pgfqpoint{3.570570in}{3.601037in}}%
\pgfpathlineto{\pgfqpoint{3.570570in}{3.564090in}}%
\pgfusepath{stroke}%
\end{pgfscope}%
\begin{pgfscope}%
\pgfpathrectangle{\pgfqpoint{0.881528in}{0.582778in}}{\pgfqpoint{5.333472in}{4.032222in}}%
\pgfusepath{clip}%
\pgfsetbuttcap%
\pgfsetroundjoin%
\pgfsetlinewidth{1.003750pt}%
\definecolor{currentstroke}{rgb}{0.000000,0.000000,0.000000}%
\pgfsetstrokecolor{currentstroke}%
\pgfsetdash{{3.700000pt}{1.600000pt}}{0.000000pt}%
\pgfpathmoveto{\pgfqpoint{3.992664in}{3.564090in}}%
\pgfpathlineto{\pgfqpoint{3.992664in}{3.520326in}}%
\pgfusepath{stroke}%
\end{pgfscope}%
\begin{pgfscope}%
\pgfpathrectangle{\pgfqpoint{0.881528in}{0.582778in}}{\pgfqpoint{5.333472in}{4.032222in}}%
\pgfusepath{clip}%
\pgfsetbuttcap%
\pgfsetroundjoin%
\pgfsetlinewidth{1.003750pt}%
\definecolor{currentstroke}{rgb}{0.000000,0.000000,0.000000}%
\pgfsetstrokecolor{currentstroke}%
\pgfsetdash{{3.700000pt}{1.600000pt}}{0.000000pt}%
\pgfpathmoveto{\pgfqpoint{3.992664in}{3.764639in}}%
\pgfpathlineto{\pgfqpoint{3.992664in}{4.117475in}}%
\pgfusepath{stroke}%
\end{pgfscope}%
\begin{pgfscope}%
\pgfpathrectangle{\pgfqpoint{0.881528in}{0.582778in}}{\pgfqpoint{5.333472in}{4.032222in}}%
\pgfusepath{clip}%
\pgfsetbuttcap%
\pgfsetroundjoin%
\pgfsetlinewidth{1.003750pt}%
\definecolor{currentstroke}{rgb}{0.000000,0.000000,0.000000}%
\pgfsetstrokecolor{currentstroke}%
\pgfsetdash{{3.700000pt}{1.600000pt}}{0.000000pt}%
\pgfpathmoveto{\pgfqpoint{3.781617in}{3.520326in}}%
\pgfpathlineto{\pgfqpoint{4.203711in}{3.520326in}}%
\pgfusepath{stroke}%
\end{pgfscope}%
\begin{pgfscope}%
\pgfpathrectangle{\pgfqpoint{0.881528in}{0.582778in}}{\pgfqpoint{5.333472in}{4.032222in}}%
\pgfusepath{clip}%
\pgfsetbuttcap%
\pgfsetroundjoin%
\pgfsetlinewidth{1.003750pt}%
\definecolor{currentstroke}{rgb}{0.000000,0.000000,0.000000}%
\pgfsetstrokecolor{currentstroke}%
\pgfsetdash{{3.700000pt}{1.600000pt}}{0.000000pt}%
\pgfpathmoveto{\pgfqpoint{3.781617in}{4.117475in}}%
\pgfpathlineto{\pgfqpoint{4.203711in}{4.117475in}}%
\pgfusepath{stroke}%
\end{pgfscope}%
\begin{pgfscope}%
\pgfpathrectangle{\pgfqpoint{0.881528in}{0.582778in}}{\pgfqpoint{5.333472in}{4.032222in}}%
\pgfusepath{clip}%
\pgfsetbuttcap%
\pgfsetroundjoin%
\definecolor{currentfill}{rgb}{0.000000,0.000000,0.000000}%
\pgfsetfillcolor{currentfill}%
\pgfsetfillopacity{0.000000}%
\pgfsetlinewidth{1.003750pt}%
\definecolor{currentstroke}{rgb}{0.000000,0.000000,0.000000}%
\pgfsetstrokecolor{currentstroke}%
\pgfsetdash{}{0pt}%
\pgfsys@defobject{currentmarker}{\pgfqpoint{-0.041667in}{-0.041667in}}{\pgfqpoint{0.041667in}{0.041667in}}{%
\pgfpathmoveto{\pgfqpoint{-0.041667in}{0.000000in}}%
\pgfpathlineto{\pgfqpoint{0.041667in}{0.000000in}}%
\pgfpathmoveto{\pgfqpoint{0.000000in}{-0.041667in}}%
\pgfpathlineto{\pgfqpoint{0.000000in}{0.041667in}}%
\pgfusepath{stroke,fill}%
}%
\begin{pgfscope}%
\pgfsys@transformshift{3.992664in}{3.515963in}%
\pgfsys@useobject{currentmarker}{}%
\end{pgfscope}%
\begin{pgfscope}%
\pgfsys@transformshift{3.992664in}{3.490332in}%
\pgfsys@useobject{currentmarker}{}%
\end{pgfscope}%
\begin{pgfscope}%
\pgfsys@transformshift{3.992664in}{4.171464in}%
\pgfsys@useobject{currentmarker}{}%
\end{pgfscope}%
\begin{pgfscope}%
\pgfsys@transformshift{3.992664in}{4.561928in}%
\pgfsys@useobject{currentmarker}{}%
\end{pgfscope}%
\end{pgfscope}%
\begin{pgfscope}%
\pgfpathrectangle{\pgfqpoint{0.881528in}{0.582778in}}{\pgfqpoint{5.333472in}{4.032222in}}%
\pgfusepath{clip}%
\pgfsetbuttcap%
\pgfsetroundjoin%
\pgfsetlinewidth{1.003750pt}%
\definecolor{currentstroke}{rgb}{0.000000,0.000000,0.000000}%
\pgfsetstrokecolor{currentstroke}%
\pgfsetdash{{3.700000pt}{1.600000pt}}{0.000000pt}%
\pgfpathmoveto{\pgfqpoint{5.347807in}{3.481334in}}%
\pgfpathlineto{\pgfqpoint{6.191995in}{3.481334in}}%
\pgfpathlineto{\pgfqpoint{6.191995in}{3.504511in}}%
\pgfpathlineto{\pgfqpoint{5.980948in}{3.527961in}}%
\pgfpathlineto{\pgfqpoint{6.191995in}{3.572679in}}%
\pgfpathlineto{\pgfqpoint{6.191995in}{3.634711in}}%
\pgfpathlineto{\pgfqpoint{5.347807in}{3.634711in}}%
\pgfpathlineto{\pgfqpoint{5.347807in}{3.572679in}}%
\pgfpathlineto{\pgfqpoint{5.558854in}{3.527961in}}%
\pgfpathlineto{\pgfqpoint{5.347807in}{3.504511in}}%
\pgfpathlineto{\pgfqpoint{5.347807in}{3.481334in}}%
\pgfusepath{stroke}%
\end{pgfscope}%
\begin{pgfscope}%
\pgfpathrectangle{\pgfqpoint{0.881528in}{0.582778in}}{\pgfqpoint{5.333472in}{4.032222in}}%
\pgfusepath{clip}%
\pgfsetbuttcap%
\pgfsetroundjoin%
\pgfsetlinewidth{1.003750pt}%
\definecolor{currentstroke}{rgb}{0.000000,0.000000,0.000000}%
\pgfsetstrokecolor{currentstroke}%
\pgfsetdash{{3.700000pt}{1.600000pt}}{0.000000pt}%
\pgfpathmoveto{\pgfqpoint{5.769901in}{3.481334in}}%
\pgfpathlineto{\pgfqpoint{5.769901in}{3.287465in}}%
\pgfusepath{stroke}%
\end{pgfscope}%
\begin{pgfscope}%
\pgfpathrectangle{\pgfqpoint{0.881528in}{0.582778in}}{\pgfqpoint{5.333472in}{4.032222in}}%
\pgfusepath{clip}%
\pgfsetbuttcap%
\pgfsetroundjoin%
\pgfsetlinewidth{1.003750pt}%
\definecolor{currentstroke}{rgb}{0.000000,0.000000,0.000000}%
\pgfsetstrokecolor{currentstroke}%
\pgfsetdash{{3.700000pt}{1.600000pt}}{0.000000pt}%
\pgfpathmoveto{\pgfqpoint{5.769901in}{3.634711in}}%
\pgfpathlineto{\pgfqpoint{5.769901in}{3.867163in}}%
\pgfusepath{stroke}%
\end{pgfscope}%
\begin{pgfscope}%
\pgfpathrectangle{\pgfqpoint{0.881528in}{0.582778in}}{\pgfqpoint{5.333472in}{4.032222in}}%
\pgfusepath{clip}%
\pgfsetbuttcap%
\pgfsetroundjoin%
\pgfsetlinewidth{1.003750pt}%
\definecolor{currentstroke}{rgb}{0.000000,0.000000,0.000000}%
\pgfsetstrokecolor{currentstroke}%
\pgfsetdash{{3.700000pt}{1.600000pt}}{0.000000pt}%
\pgfpathmoveto{\pgfqpoint{5.558854in}{3.287465in}}%
\pgfpathlineto{\pgfqpoint{5.980948in}{3.287465in}}%
\pgfusepath{stroke}%
\end{pgfscope}%
\begin{pgfscope}%
\pgfpathrectangle{\pgfqpoint{0.881528in}{0.582778in}}{\pgfqpoint{5.333472in}{4.032222in}}%
\pgfusepath{clip}%
\pgfsetbuttcap%
\pgfsetroundjoin%
\pgfsetlinewidth{1.003750pt}%
\definecolor{currentstroke}{rgb}{0.000000,0.000000,0.000000}%
\pgfsetstrokecolor{currentstroke}%
\pgfsetdash{{3.700000pt}{1.600000pt}}{0.000000pt}%
\pgfpathmoveto{\pgfqpoint{5.558854in}{3.867163in}}%
\pgfpathlineto{\pgfqpoint{5.980948in}{3.867163in}}%
\pgfusepath{stroke}%
\end{pgfscope}%
\begin{pgfscope}%
\pgfpathrectangle{\pgfqpoint{0.881528in}{0.582778in}}{\pgfqpoint{5.333472in}{4.032222in}}%
\pgfusepath{clip}%
\pgfsetbuttcap%
\pgfsetroundjoin%
\definecolor{currentfill}{rgb}{0.000000,0.000000,0.000000}%
\pgfsetfillcolor{currentfill}%
\pgfsetfillopacity{0.000000}%
\pgfsetlinewidth{1.003750pt}%
\definecolor{currentstroke}{rgb}{0.000000,0.000000,0.000000}%
\pgfsetstrokecolor{currentstroke}%
\pgfsetdash{}{0pt}%
\pgfsys@defobject{currentmarker}{\pgfqpoint{-0.041667in}{-0.041667in}}{\pgfqpoint{0.041667in}{0.041667in}}{%
\pgfpathmoveto{\pgfqpoint{-0.041667in}{0.000000in}}%
\pgfpathlineto{\pgfqpoint{0.041667in}{0.000000in}}%
\pgfpathmoveto{\pgfqpoint{0.000000in}{-0.041667in}}%
\pgfpathlineto{\pgfqpoint{0.000000in}{0.041667in}}%
\pgfusepath{stroke,fill}%
}%
\begin{pgfscope}%
\pgfsys@transformshift{5.769901in}{3.271105in}%
\pgfsys@useobject{currentmarker}{}%
\end{pgfscope}%
\begin{pgfscope}%
\pgfsys@transformshift{5.769901in}{3.229659in}%
\pgfsys@useobject{currentmarker}{}%
\end{pgfscope}%
\begin{pgfscope}%
\pgfsys@transformshift{5.769901in}{4.511211in}%
\pgfsys@useobject{currentmarker}{}%
\end{pgfscope}%
\begin{pgfscope}%
\pgfsys@transformshift{5.769901in}{4.077665in}%
\pgfsys@useobject{currentmarker}{}%
\end{pgfscope}%
\end{pgfscope}%
\begin{pgfscope}%
\pgfpathrectangle{\pgfqpoint{0.881528in}{0.582778in}}{\pgfqpoint{5.333472in}{4.032222in}}%
\pgfusepath{clip}%
\pgfsetrectcap%
\pgfsetroundjoin%
\pgfsetlinewidth{1.505625pt}%
\definecolor{currentstroke}{rgb}{0.121569,0.466667,0.705882}%
\pgfsetstrokecolor{currentstroke}%
\pgfsetdash{}{0pt}%
\pgfpathmoveto{\pgfqpoint{2.659736in}{0.582778in}}%
\pgfpathlineto{\pgfqpoint{2.659736in}{4.615000in}}%
\pgfusepath{stroke}%
\end{pgfscope}%
\begin{pgfscope}%
\pgfpathrectangle{\pgfqpoint{0.881528in}{0.582778in}}{\pgfqpoint{5.333472in}{4.032222in}}%
\pgfusepath{clip}%
\pgfsetrectcap%
\pgfsetroundjoin%
\pgfsetlinewidth{1.505625pt}%
\definecolor{currentstroke}{rgb}{0.121569,0.466667,0.705882}%
\pgfsetstrokecolor{currentstroke}%
\pgfsetdash{}{0pt}%
\pgfpathmoveto{\pgfqpoint{4.436973in}{0.582778in}}%
\pgfpathlineto{\pgfqpoint{4.436973in}{4.615000in}}%
\pgfusepath{stroke}%
\end{pgfscope}%
\begin{pgfscope}%
\pgfpathrectangle{\pgfqpoint{0.881528in}{0.582778in}}{\pgfqpoint{5.333472in}{4.032222in}}%
\pgfusepath{clip}%
\pgfsetrectcap%
\pgfsetroundjoin%
\pgfsetlinewidth{1.003750pt}%
\definecolor{currentstroke}{rgb}{0.000000,0.000000,0.000000}%
\pgfsetstrokecolor{currentstroke}%
\pgfsetdash{}{0pt}%
\pgfpathmoveto{\pgfqpoint{1.115761in}{1.042295in}}%
\pgfpathlineto{\pgfqpoint{1.537855in}{1.042295in}}%
\pgfusepath{stroke}%
\end{pgfscope}%
\begin{pgfscope}%
\pgfpathrectangle{\pgfqpoint{0.881528in}{0.582778in}}{\pgfqpoint{5.333472in}{4.032222in}}%
\pgfusepath{clip}%
\pgfsetrectcap%
\pgfsetroundjoin%
\pgfsetlinewidth{1.003750pt}%
\definecolor{currentstroke}{rgb}{0.000000,0.000000,0.000000}%
\pgfsetstrokecolor{currentstroke}%
\pgfsetdash{}{0pt}%
\pgfpathmoveto{\pgfqpoint{2.892998in}{1.223349in}}%
\pgfpathlineto{\pgfqpoint{3.315092in}{1.223349in}}%
\pgfusepath{stroke}%
\end{pgfscope}%
\begin{pgfscope}%
\pgfpathrectangle{\pgfqpoint{0.881528in}{0.582778in}}{\pgfqpoint{5.333472in}{4.032222in}}%
\pgfusepath{clip}%
\pgfsetrectcap%
\pgfsetroundjoin%
\pgfsetlinewidth{1.003750pt}%
\definecolor{currentstroke}{rgb}{0.000000,0.000000,0.000000}%
\pgfsetstrokecolor{currentstroke}%
\pgfsetdash{}{0pt}%
\pgfpathmoveto{\pgfqpoint{4.670236in}{1.385588in}}%
\pgfpathlineto{\pgfqpoint{5.092329in}{1.385588in}}%
\pgfusepath{stroke}%
\end{pgfscope}%
\begin{pgfscope}%
\pgfpathrectangle{\pgfqpoint{0.881528in}{0.582778in}}{\pgfqpoint{5.333472in}{4.032222in}}%
\pgfusepath{clip}%
\pgfsetbuttcap%
\pgfsetroundjoin%
\pgfsetlinewidth{1.003750pt}%
\definecolor{currentstroke}{rgb}{0.000000,0.000000,0.000000}%
\pgfsetstrokecolor{currentstroke}%
\pgfsetdash{{3.700000pt}{1.600000pt}}{0.000000pt}%
\pgfpathmoveto{\pgfqpoint{2.004380in}{1.056747in}}%
\pgfpathlineto{\pgfqpoint{2.426474in}{1.056747in}}%
\pgfusepath{stroke}%
\end{pgfscope}%
\begin{pgfscope}%
\pgfpathrectangle{\pgfqpoint{0.881528in}{0.582778in}}{\pgfqpoint{5.333472in}{4.032222in}}%
\pgfusepath{clip}%
\pgfsetbuttcap%
\pgfsetroundjoin%
\pgfsetlinewidth{1.003750pt}%
\definecolor{currentstroke}{rgb}{0.000000,0.000000,0.000000}%
\pgfsetstrokecolor{currentstroke}%
\pgfsetdash{{3.700000pt}{1.600000pt}}{0.000000pt}%
\pgfpathmoveto{\pgfqpoint{3.781617in}{3.659934in}}%
\pgfpathlineto{\pgfqpoint{4.203711in}{3.659934in}}%
\pgfusepath{stroke}%
\end{pgfscope}%
\begin{pgfscope}%
\pgfpathrectangle{\pgfqpoint{0.881528in}{0.582778in}}{\pgfqpoint{5.333472in}{4.032222in}}%
\pgfusepath{clip}%
\pgfsetbuttcap%
\pgfsetroundjoin%
\pgfsetlinewidth{1.003750pt}%
\definecolor{currentstroke}{rgb}{0.000000,0.000000,0.000000}%
\pgfsetstrokecolor{currentstroke}%
\pgfsetdash{{3.700000pt}{1.600000pt}}{0.000000pt}%
\pgfpathmoveto{\pgfqpoint{5.558854in}{3.527961in}}%
\pgfpathlineto{\pgfqpoint{5.980948in}{3.527961in}}%
\pgfusepath{stroke}%
\end{pgfscope}%
\begin{pgfscope}%
\pgfsetrectcap%
\pgfsetmiterjoin%
\pgfsetlinewidth{0.803000pt}%
\definecolor{currentstroke}{rgb}{0.000000,0.000000,0.000000}%
\pgfsetstrokecolor{currentstroke}%
\pgfsetdash{}{0pt}%
\pgfpathmoveto{\pgfqpoint{0.881528in}{0.582778in}}%
\pgfpathlineto{\pgfqpoint{0.881528in}{4.615000in}}%
\pgfusepath{stroke}%
\end{pgfscope}%
\begin{pgfscope}%
\pgfsetrectcap%
\pgfsetmiterjoin%
\pgfsetlinewidth{0.803000pt}%
\definecolor{currentstroke}{rgb}{0.000000,0.000000,0.000000}%
\pgfsetstrokecolor{currentstroke}%
\pgfsetdash{}{0pt}%
\pgfpathmoveto{\pgfqpoint{6.215000in}{0.582778in}}%
\pgfpathlineto{\pgfqpoint{6.215000in}{4.615000in}}%
\pgfusepath{stroke}%
\end{pgfscope}%
\begin{pgfscope}%
\pgfsetrectcap%
\pgfsetmiterjoin%
\pgfsetlinewidth{0.803000pt}%
\definecolor{currentstroke}{rgb}{0.000000,0.000000,0.000000}%
\pgfsetstrokecolor{currentstroke}%
\pgfsetdash{}{0pt}%
\pgfpathmoveto{\pgfqpoint{0.881528in}{0.582778in}}%
\pgfpathlineto{\pgfqpoint{6.215000in}{0.582778in}}%
\pgfusepath{stroke}%
\end{pgfscope}%
\begin{pgfscope}%
\pgfsetrectcap%
\pgfsetmiterjoin%
\pgfsetlinewidth{0.803000pt}%
\definecolor{currentstroke}{rgb}{0.000000,0.000000,0.000000}%
\pgfsetstrokecolor{currentstroke}%
\pgfsetdash{}{0pt}%
\pgfpathmoveto{\pgfqpoint{0.881528in}{4.615000in}}%
\pgfpathlineto{\pgfqpoint{6.215000in}{4.615000in}}%
\pgfusepath{stroke}%
\end{pgfscope}%
\begin{pgfscope}%
\pgftext[x=1.859331in,y=0.744067in,,base]{\sffamily\fontsize{10.000000}{12.000000}\selectfont p = 5.0e-01}%
\end{pgfscope}%
\begin{pgfscope}%
\pgftext[x=3.637155in,y=0.744067in,,base]{\sffamily\fontsize{10.000000}{12.000000}\selectfont p = 8.5e-61}%
\end{pgfscope}%
\begin{pgfscope}%
\pgftext[x=5.414979in,y=0.744067in,,base]{\sffamily\fontsize{10.000000}{12.000000}\selectfont p = 8.0e-53}%
\end{pgfscope}%
\begin{pgfscope}%
\pgfsetbuttcap%
\pgfsetmiterjoin%
\definecolor{currentfill}{rgb}{1.000000,1.000000,1.000000}%
\pgfsetfillcolor{currentfill}%
\pgfsetfillopacity{0.800000}%
\pgfsetlinewidth{1.003750pt}%
\definecolor{currentstroke}{rgb}{0.800000,0.800000,0.800000}%
\pgfsetstrokecolor{currentstroke}%
\pgfsetstrokeopacity{0.800000}%
\pgfsetdash{}{0pt}%
\pgfpathmoveto{\pgfqpoint{0.978750in}{4.096174in}}%
\pgfpathlineto{\pgfqpoint{2.469134in}{4.096174in}}%
\pgfpathquadraticcurveto{\pgfqpoint{2.496911in}{4.096174in}}{\pgfqpoint{2.496911in}{4.123952in}}%
\pgfpathlineto{\pgfqpoint{2.496911in}{4.517778in}}%
\pgfpathquadraticcurveto{\pgfqpoint{2.496911in}{4.545556in}}{\pgfqpoint{2.469134in}{4.545556in}}%
\pgfpathlineto{\pgfqpoint{0.978750in}{4.545556in}}%
\pgfpathquadraticcurveto{\pgfqpoint{0.950972in}{4.545556in}}{\pgfqpoint{0.950972in}{4.517778in}}%
\pgfpathlineto{\pgfqpoint{0.950972in}{4.123952in}}%
\pgfpathquadraticcurveto{\pgfqpoint{0.950972in}{4.096174in}}{\pgfqpoint{0.978750in}{4.096174in}}%
\pgfpathclose%
\pgfusepath{stroke,fill}%
\end{pgfscope}%
\begin{pgfscope}%
\pgfsetrectcap%
\pgfsetroundjoin%
\pgfsetlinewidth{1.505625pt}%
\definecolor{currentstroke}{rgb}{0.000000,0.000000,0.000000}%
\pgfsetstrokecolor{currentstroke}%
\pgfsetdash{}{0pt}%
\pgfpathmoveto{\pgfqpoint{1.006528in}{4.433088in}}%
\pgfpathlineto{\pgfqpoint{1.284306in}{4.433088in}}%
\pgfusepath{stroke}%
\end{pgfscope}%
\begin{pgfscope}%
\pgftext[x=1.395417in,y=4.384477in,left,base]{\sffamily\fontsize{10.000000}{12.000000}\selectfont ChromeNoQuic}%
\end{pgfscope}%
\begin{pgfscope}%
\pgfsetbuttcap%
\pgfsetroundjoin%
\pgfsetlinewidth{1.505625pt}%
\definecolor{currentstroke}{rgb}{0.000000,0.000000,0.000000}%
\pgfsetstrokecolor{currentstroke}%
\pgfsetdash{{5.550000pt}{2.400000pt}}{0.000000pt}%
\pgfpathmoveto{\pgfqpoint{1.006528in}{4.229231in}}%
\pgfpathlineto{\pgfqpoint{1.284306in}{4.229231in}}%
\pgfusepath{stroke}%
\end{pgfscope}%
\begin{pgfscope}%
\pgftext[x=1.395417in,y=4.180620in,left,base]{\sffamily\fontsize{10.000000}{12.000000}\selectfont ChromeQuic}%
\end{pgfscope}%
\begin{pgfscope}%
\pgfline{\pgfpoint{3cm}{6cm}}{\pgfpoint{4cm}{6cm}}
\pgfline{\pgfpoint{3cm}{10cm}}{\pgfpoint{4cm}{10cm}}
\pgfline{\pgfpoint{3.5cm}{10cm}}{\pgfpoint{3.5cm}{9.5cm}}
\pgfline{\pgfpoint{3.5cm}{6cm}}{\pgfpoint{3.5cm}{6.5cm}}
\pgfline{\pgfpoint{2.5cm}{9.5cm}}{\pgfpoint{4.5cm}{9.5cm}}
\pgfline{\pgfpoint{2.5cm}{6.5cm}}{\pgfpoint{4.5cm}{6.5cm}}
\pgfline{\pgfpoint{2.5cm}{6.5cm}}{\pgfpoint{2.5cm}{7.5cm}}
\pgfline{\pgfpoint{4.5cm}{6.5cm}}{\pgfpoint{4.5cm}{7.5cm}}
\pgfline{\pgfpoint{2.5cm}{8.5cm}}{\pgfpoint{2.5cm}{9.5cm}}
\pgfline{\pgfpoint{4.5cm}{8.5cm}}{\pgfpoint{4.5cm}{9.5cm}}
\pgfline{\pgfpoint{4.5cm}{8.5cm}}{\pgfpoint{4cm}{8cm}}
\pgfline{\pgfpoint{4cm}{8cm}}{\pgfpoint{4.5cm}{7.5cm}}
\pgfline{\pgfpoint{2.5cm}{8.5cm}}{\pgfpoint{3cm}{8cm}}
\pgfline{\pgfpoint{3cm}{8cm}}{\pgfpoint{2.5cm}{7.5cm}}
\pgfline{\pgfpoint{3cm}{8cm}}{\pgfpoint{4cm}{8cm}}
\end{pgfscope}%
\begin{pgfscope}%
\pgfsetendarrow{\pgfarrowtriangle{4pt}}
\pgfline{\pgfpoint{6cm}{6cm}}{\pgfpoint{4.2cm}{6cm}}
\pgfline{\pgfpoint{6cm}{6.5cm}}{\pgfpoint{4.7cm}{6.5cm}}
\pgfline{\pgfpoint{6cm}{8cm}}{\pgfpoint{4.2cm}{8cm}}
\pgfline{\pgfpoint{6cm}{9.5cm}}{\pgfpoint{4.7cm}{9.5cm}}
\pgfline{\pgfpoint{6cm}{10cm}}{\pgfpoint{4.2cm}{10cm}}
\pgfputat{\pgfpoint{5cm}{5.5cm}}{\pgfbox[left,center]{Percentile}}
\pgfputat{\pgfpoint{6cm}{6cm}}{\pgfbox[left,center]{5th}}
\pgfputat{\pgfpoint{6cm}{6.5cm}}{\pgfbox[left,center]{25th}}
\pgfputat{\pgfpoint{6cm}{8cm}}{\pgfbox[left,center]{50th}}
\pgfputat{\pgfpoint{6cm}{9.5cm}}{\pgfbox[left,center]{75th}}
\pgfputat{\pgfpoint{6cm}{10cm}}{\pgfbox[left,center]{95th}}
\pgfputat{\pgfpoint{5cm}{9cm}}{\pgfbox[left,center]{\scriptsize{Median 95\%}}}
\pgfputat{\pgfpoint{5cm}{8.75cm}}{\pgfbox[left,center]{\scriptsize{confidence}}}
\pgfputat{\pgfpoint{5cm}{8.5cm}}{\pgfbox[left,center]{\scriptsize{interval}}}
\pgfsetendarrow{\pgfarrowto}
\pgfline{\pgfpoint{5cm}{8.75cm}}{\pgfpoint{4.55cm}{7.5cm}}
\pgfline{\pgfpoint{5cm}{8.75cm}}{\pgfpoint{4.55cm}{8.5cm}}
\end{pgfscope}%
\end{pgfpicture}%
\makeatother%
\endgroup%

%% file: related_work.tex
\section{Contribution and related work}

Performance comparisons of GQUIC and TCP have already been conducted under various network conditions and for various applications~\cite{long-look, tcp-vs-quic}. For instance, the authors in~\cite{quic-satcom} have performed an evaluation of GQUIC on an emulated platform with scenarios involving SATCOM in LEO and GEO contexts. They concluded that GQUIC outperforms TCP but their testbed did not include any PEP. On the contrary, our study demonstrates that transparent proxying is the cornerstone of better Page Load Time with TCP when compared to GQUIC.

The impossibility for GQUIC to benefit from the PEP technology has already been identified in~\cite{long-look}. Here again, the authors concluded that GQUIC continues to outperform TCP even when the later is split by a proxy. Our tests on a real access highlight the influence of complex PEP deployment schemes on the comparative performance of both protocols.

This paper completes the related work by reporting the first evaluations of GQUIC using a real public SATCOM access. We consider two web-pages: one picture with a $5.3$\,MB total size and one Google's 404 page with two objects and $11$\,kB of total size. The main conclusions are the following: 
\begin{itemize}
        \item for a large web page, the page load time is approximately twice longer with GQUIC compared to TCP;
        \item this difference in larger page load time resides in the poor performance of the non-delegated Congestion Controller in GQUIC;
        \item although faster, GQUIC connection establishment does not compensate the above issue.
\end{itemize}

As it is, GQUIC may result in poor end user experience on a high BDP network. However, inducing specific tuning for SATCOM networks in the standards and in actual deployments may be complicated.

%% file: discussion.tex
\section{Discussion}
\label{sec:discussion}

The important variability in applications' requirements makes it hard to
define a transport protocol that suits them all.  The relevance of
application and transport layers protocols depends on both (a) how
application data packets are generated and carried out and (b) the network
underneath. Taking as example a SATCOM Internet access, the performance
of different web applications highly depends on the web pages
characteristics~\cite{perfo-web-sat}. 
There is no ``transport layer silver
bullet'', \textit{i.e.} a transport protocol that would suit to any
application and any network conditions~\cite{no-silver-bullet}.

\subsection{ISP point-of-view}
\label{sec:discussion:network-operator}

For a SATCOM ISP, the deployment of any-QUIC could be seen as interesting:
\begin{itemize}
\item we can expect interesting gains for short flows,
\item we would not need PEP that are expensive and sometimes complicated to maintain and operate. 
\end{itemize}
However, the performance gains for large files
transmissions, the inadequacy of the end-to-end congestion control to
SATCOM links and the rapid evolutivity of the protocol may be seen as a
threat to good end-user quality of experience.

Moreover, such as illustrated by the discussions on the spin-bit at the IETF, ISP 
could encounter issues in identifying where the loss occurs in the network.

\subsection{Towards a middlebox-friendly any-QUIC}
\label{sec:discussion:friendly-quic}

To date, GQUIC does not seem to be blocked by ISP companies, according to~\cite{quic-in-the-wild} and our experiments. 
It may not be the case when this UDP
traffic becomes a greater proportion of the total traffic. Indeed, for the reasons
mentioned in~\ref{sec:discussion:network-operator}, quickly evolving protocols
may not be easy to deal with, from an ISP point-of-view. Interactions
between the end-to-end protocol and the operator middle-boxes could be enabled,
such as discussed in~\cite{quic-manageability}. Moreover, bits such as the spin-bit could be made
available for the ISP operations, such as load-balancing and statistics on
the current flows. 

To
adapt the congestion control and data rate transmission to a specific
SATCOM scenario, more important modifications may be required. 
They may
involve changing the advertised receive window, such as in IP-Explicit Rate Notification (IP-ERN)
architectures~\cite{ip-ern}, delegate the security to the network operator
for better quality of experiences or update the browser~\cite{web-browser-viasat}.
These specifications could be negotiated at initialization such as discussed in the transport draft of IQUIC~\cite{quic-ietf-transport}. 


